\documentclass[12pt]{iopart}

\expandafter\let\csname equation*\endcsname\relax
\expandafter\let\csname endequation*\endcsname\relax

\usepackage{graphicx}
\usepackage{xspace}
\usepackage[usenames,dvipsnames]{color}
\usepackage{amsmath,amssymb}
\usepackage{mathrsfs}
\usepackage{bm}
\usepackage[normalem]{ulem} 
\usepackage{listings}
\usepackage{tensor}
\usepackage{cite}

\usepackage[utf8]{inputenc}

\usepackage[colorlinks=true,citecolor=blue,urlcolor=blue]{hyperref}

\newcommand{\harald}[1]{\textcolor{ForestGreen}{#1}}

\begin{document}

\title[GPU-Accelerated Simulations of Isolated Black Holes]{
GPU-Accelerated Simulations of Isolated Black Holes
}

\author{Adam G.~M.~Lewis${}^{1,2,3}$} 
\address{${}^1$Canadian Institute for Theoretical Astrophysics, 60 St. George St., Toronto, M5S 3H8, Ontario, Canada}
\address{${}^2$Department of Physics, University of Toronto, 60 St. George St., Toronto, M5S 1A7, Ontario, Canada}
\address{
  ${}^3$Perimeter Institute for Theoretical Physics,
  31 Caroline St.~N, Waterloo, ON N2L 2Y5, Canada} 
\ead{alewis@perimeterinstitute.ca}

\author{Harald P. Pfeiffer${}^{1,4}$}

\address{${}^4$Max-Planck-Institut f\"{u}r Gravitationsphysik (Albert-Einstein-Institut), Wissenschaftspark Potsdam-Golm, Am M\"{u}hlenberg 1, 14476, Potsdam, Germany}
\ead{harald.pfeiffer@aei.mpg.de}

\date{\today}

\begin{abstract}
  We present a port of the numerical relativity code \texttt{SpEC} which is
  capable of running on NVIDIA GPUs. Since this code must be maintained in
  parallel with \texttt{SpEC} itself, a primary design consideration is to
  perform as few explicit code changes as possible. We therefore rely on a 
  hierarchy of automated porting strategies. At the highest level we use 
  \texttt{TLoops}, a C++ library of our design, to automatically emit
  CUDA code equivalent to tensorial expressions written into C++ source using
  a syntax similar to analytic calculation. Next, we trace out
  and cache explicit matrix representations of the numerous linear transformations
  in the \texttt{SpEC} code, which allows these to be performed on the GPU using
  pre-existing matrix-multiplication libraries. We port the few remaining important
  modules by hand. In this paper we detail the specifics of our port, and 
  present benchmarks of it simulating isolated black hole spacetimes on
  several generations of NVIDIA GPU.
\end{abstract}

%\pacs{}

\maketitle

%%%%%%%%%%%%%%%%%%%%%%%%%%%%%%%%%%%%%%%%%%%%%%%%%%%%%%%%%%%%%%%%%%%%%%%%%%%%%%%
\section{Introduction}
%%%%%%%%%%%%%%%%%%%%%%%%%%%%%%%%%%%%%%%%%%%%%%%%%%%%%%%%%%%%%%%%%%%%%%%%%%%%%%%

Numerical relativity (NR), the direct numerical integration of the Einstein
field equations, is now a mature subfield of computational physics,
with stable binary black hole evolutions possible since 2005
\cite{pretorius2005evolution, pretorius2005numerical, campanelli2006accurate, Baker2006a, lastorbit2006}.
Binary black hole NR is of central importance for gravitational waveform modeling. 
For instance, the \texttt{SpEC} waveform catalogues \cite{Mroue:2013xna, Chu:2015kft} were used in the development
of EOB-waveform models \cite{Bohe:2016gbl, Pan:2013rra, Taracchini:2013rva}. Waveform
models calibrated to numerical relativity were used to analyse the resent BBH gravitational wave events 
detected by LIGO and Virgo~\cite{LIGOVirgo2016a, TheLIGOScientific:2016qqj}. 
Furthermore, NR binary black hole simulations were used to assess systematic errors of parameter estimation
of these GW events \cite{Abbott:2016apu, Abbott:2016wiq, Lovelace:2016uwp}. 

Detailed knowledge of expected waveforms, themselves coming ultimately
from NR simulations, are required by these detectors to maximize sensitivities, 
to interpret observation, 
and to make tests of general relativity~\cite{LIGO-DataAnalysis-Whitepaper:2015}.  
Ground based detectors' relative
insensitivity to e.g. eccentric binaries is, conversely, in part due to a lack of
production-quality simulations in the eccentric region of parameter space \cite{Lewis:2016lgx}, 
a situation which also impairs comparisons with analytic theory.

The intricacy of the Einstein equations presents two 
challenges to NR. First, interesting simulations are 
expensive, with wallclock times measured in weeks or months. 
Second, codes able to perform such simulations are quite intricate
from a software engineering perspective. Applying them to new regions of
the binary black hole parameter space - let alone to new
spacetimes - can require months of effort by small groups of 
experts. These issues are difficult to address simultaneously,
since improving runtime tends to complicate code, and vice versa.

Twenty years ago, the simplest solution would have been to 
simply wait for hardware to
improve. Unfortunately CPU clock frequencies have been essentially 
static for some time now, 
with new high-performance computers instead employing increasingly massive levels of 
parallelism. But few codes scale to $100 000$s of CPUs 
without considerable reformulation.

Easier speedups can sometimes be achieved using ``accelerated" architectures.
A popular choice is the
Graphics Processing Unit (GPU). Problems in graphical computation might involve,
for example, computing pixel states as a virtual object moves three-dimensionally.
Each pixel is data independent, and the fundamental operations are linear
transformations such as rotations. Thus, viewed more abstractly than perhaps 
originally intended, GPUs are optimized for
highly parallelizable linear operations. These are performed by slow, but numerous
and tightly-coupled, processors connected by various
hierarchies of memory. The tight coupling and fast RAM make intra-GPU
communication inexpensive. Because of this, GPUs are, for suitable problems, 
potentially dramatically superior to CPUs in terms of e.g. FLOPS-per-watt. 

The extra parallel cores replace the CPU's extensive control circuitry,
which for example rearranges instructions to optimize single-thread performance,
along with much of its cache memory, which increases the speed of non-contiguous memory access.
GPUs are thus less flexible than CPUs, and less able to handle 
fundamentally serial problems. But even for less-than-ideal use-cases, GPUs
have a major
advantage over CPUs: they continue to demonstrate Moore's-law-like FLOPS/year
scaling with new releases. Therefore, once the initial investment 
is made to produce a port, further speedups can be made 
by simply buying new hardware\footnote{The speedup-over-time is 
largely due to newer cards supporting yet-higher levels of parallelism. 
Problems that already exhaust the parallelism of an existing card will 
not benefit from the scaling, or will benefit only weakly.}.

GPUs now enjoy widespread and increasing use
as ``accelerators" of numerically-intensive, linear-algebra-heavy tasks such 
as physical simulation and deep machine-learning. At the time of writing, they
have not been widely adopted in NR, due likely to the complexity introduced
by the low-level nature of GPU coding. 
Some previous applications of GPU computing to NR do, however, exist.
Zink \cite{Zink:2008} used a CUDA-based finite-difference code to evolve a gauge wave
upon Minkowski spacetime. Later, he developed HORIZON
\cite{Zink:2011}, a GPU-accelerated GRMHD code. Yang et al. \cite{Yang:2016} simulated
plunging black-hole binaries in the BSSN formalism using a finite-difference 
CUDA port of the numerical relativity code AMSS-NCKU \cite{PhysRevD.78.124011, PhysRevD.82.024005}in concert with an AMR-like algorithm of their own design.
Chen \cite{Chen:2015} used a GPU-accelerated approach to solve 
sample coupled elliptic equations using the spectral collocation and spectral 
Galerkin methods. The Teukolsky master equations \cite{1973ApJ...185...635T}
describing perturbations to the Kerr spacetime have been solved using the Cell
processor SDK \cite{Khanna:2009me}, OpenCL \cite{choudhary2010exploration, Khanna:2010jv}, 
and CUDA \cite{Khanna:2013}. Herrmann et. al \cite{Herrmann_blackhole} used CUDA 
to integrate the PN equations at an unspecified truncation order. Brugmann 
\cite{BRUGMANN2013} developed
a GPU algorithm to integrate PDE systems relating time-dependent tensor fields
on a spherical shell using pseudospectral methods.
Perhaps most similarly to our 
work, automatically-generated GPU code has been used to benchmark binary black
holes inspirals simulated using the Einstein Toolkit \cite{ChemoraPaper}.

In the present work we describe our approach to the GPU porting of a specific
NR code, the Spectral Einstein Code ($\mathtt{SpEC}$) \cite{SpEC}. 
Multi-domain spectral methods like those used by $\mathtt{SpEC}$ solve 
PDEs by dividing the simulation volume into ``domains"
upon which the solution is smooth. Within each domain, the solution can 
then be represented
as weights to a truncated sum of basis functions. 
Data thus represented are nonlocal in space. Every 
processor working on the same domain will generally require access
to the full spectrum. 

$\mathtt{SpEC}$ already uses MPI to assign each domain to a different processor. 
In setups where multiple CPU cores share RAM, it should also be possible to assign multiple
cores to a \emph{single} domain, but this has not been achieved in practice despite 
repeated efforts.
On the other hand, the many 
processors within a single GPU also share a unified pool of  
memory. Access to this memory is sufficiently fast that the entire GPU can work 
on the same domain. GPUs, in other words, enable $\mathtt{SpEC}$ to scale to higher levels of 
parallelism than would be otherwise possible. Conversely, $\mathtt{SpEC}$'s inability to
utilize OpenMP means that the relevant benchmark in this work is the performance of a
single CPU to an entire GPU. 

In principle the Einstein equations are merely a specific example of a 
hyperbolic PDE system to be solved, which presents challenges on either the CPU
or GPU identical to any other. The algorithmic reasoning employed in this
work is indeed quite straightforward, and 
our challenges have instead been more practical. In simple terms, the $\mathtt{SpEC}$ 
source code is very long and complicated. 
Computational effort is spent mostly on a relatively
small number of modules, but in practice these are each implemented
by a complex and diverse set of subclasses depending on, for example, the topology of 
the domain (cubes, cylinders, spherical shells, etc.) upon which they operate. 

Producing GPU equivalents of all the necessary
instances would involve considerable effort. More problematically, upon 
completion code maintenance would become infeasibly difficult, since
consistency between the CPU and GPU code bases would have to be continuously maintained
at each revision. Of course, only relatively little code is actually 
performance-critical enough to yield practical benefits from GPU acceleration.
Porting only these critical segments, however, results in numerous
expensive CPU-GPU memory synchronizations, since the input to and output 
from the critical modules must be accessible to the relevant processor. 
This expense swamps any speedup in practice. 

To keep the amount of redundant code manageable, we use a combination of
porting strategies relying upon various levels of automation. At the highest
level of automation, we have developed a C++ library, \texttt{TLoops},  
which provides technology to write spatially-decoupled tensor-algebraic expressions directly
into C++ source code. For example, the \texttt{TLoops} expression of listing \ref{lst:tloopscode}
yields output equivalent to that of listing \ref{lst:forloops}.
\begin{lstlisting}[caption={\label{lst:tloopscode}Example TLoops expression.}, language=C++]
  Tensor<DataMesh> dtg, K, db;
  DataMesh alpha;
  // initialize dtg, K, db, alpha
  dtg(Sym<0,1>(),i_,j_)=-2*alpha*K(i_,j_)+db(i_,j_)+db(j_,i_);
\end{lstlisting}
\begin{lstlisting}[caption={\label{lst:forloops}C-style code, equivalent to Listing \ref{lst:tloopscode}.}, language=C]
  Tensor<DataMesh> dtg, K, db;
  DataMesh alpha;
  // initialize dtg, K, db, alpha
  for(int i=0; i<3; ++i) {
    for(int j=0; j<=i; ++j) {
      for(int a=0; a<N; ++a) {
        dtg(i,j)[a]=-2.0*alpha[a]*K(i,j)[a]+db(i,j)[a]+db(j,i)[a];
      }
    }
  }
\end{lstlisting}
Here \texttt{DataMesh} is \texttt{SpEC}'s
array class, representing one double precision value at each point on a simulation
domain, while \texttt{Tensor<DataMesh>} is a container class representing 
one \texttt{DataMesh} for each component of a tensor (field). \texttt{TLoops}
can run the code in Listing \ref{lst:tloopscode} at once, or can output
equivalent GPU code to be linked against a separate \texttt{SpEC} compilation,
allowing for efficient GPU porting with very minimal effort (the replacement
of code in the form of Listing \ref{lst:forloops} with \texttt{TLoops} expressions
such as those of Listing \ref{lst:tloopscode}). \texttt{TLoops} will also
output equivalent GPU code to the line inside the loop in Listing \ref{lst:forloops},
allowing almost the entire \texttt{SpEC} code to be (inefficiently) ported 
at once. 

\texttt{TLoops}, thus, allows \texttt{SpEC} to keep data always on the
GPU without any additional coding. Code segments which consume large
amounts of wallclock time and which consist mostly of tensor manipulations,
such as the code \texttt{SpEC} uses to advance the Einstein equations 
in their generalized harmonic formulation by
a timestep, can usually be sped up by at least 10 times relative to the GPU
through the use of \texttt{TLoops} statements such as Listing \ref{lst:tloopscode}.

At the next levels of automation, there are a number of modules that
are performance-critical, but which cannot be handled by \texttt{TLoops}
because they are not spatially-decoupled (i.e. their output at a given
gridpoint depends on simulation data at other gridpoints). These modules,
which include the differentiator for example, turn out to have several
key features in common. First, each may represent any of 
various transformations, and often multiple implementations of 
each. The differentiator, for example, may be handled using a matrix
multiplication, or by spectral methods whose details depend on the domain
topology, with the choice being made by the user at runtime. The number
of possible execution branches is too large to port everything by hand.

Fortunately, all such modules used by \texttt{SpEC} turn out to represent
\emph{linear} transformations whose matrix representations have finite rank. 
Furthermore, while the 
number of \emph{possible} execution branches is very large, the number
of \emph{actual} branches encountered by a particular process is in practice
manageably small. We therefore write GPU code which implements \harald{\sout{finite}}
linear transformations, given an explicit matrix representation of them.
In
this way many lines of code may be ported with relatively little effort.

There are, finally, some modules which are neither amenable
to \texttt{TLoops} nor to the cached linear transformation approach.
This last class includes, for example, sequences of contractions with Jacobian
matrices designed to transform the spatial coordinates of a spacetime tensor
while leaving the temporal coordinates unchanged.
These are few enough to simply port by hand.

In concert, thus, our strategy consists first of automated porting of existing
expressions using \texttt{TLoops}, with the loops in some especially
important modules replaced by \texttt{TLoops} expressions. Next, we
port linear transformations by tracing out their explicit matrix
representation, and port the few remaining important segments by hand.
 
The rest of this paper is structured as follows. Section 2 describes $\mathtt{SpEC}$
and introduces the GPU architecture.  Section 3 gives a 
detailed, module-by-module description of our port including benchmarks. 
Section 4 presents and discusses holistic benchmarks of the entire SingleBH 
test case. Finally, Section 5 draws conclusions and motivates future research.

%%%%%%%%%%%%%%%%%%%%%%%%%%%%%%%%%%%%%%%%%%%%%%%%%%%%%%%%%%%%%%%%%%%%%%%%%%%%%%%
\section{Background}
%%%%%%%%%%%%%%%%%%%%%%%%%%%%%%%%%%%%%%%%%%%%%%%%%%%%%%%%%%%%%%%%%%%%%%%%%%%%%%%

\subsection{The Spectral Einstein Code}
The Spectral Einstein Code ($\mathtt{SpEC}$) \cite{SpEC} is a C++ code designed to solve Einstein’s 
equations of general relativity.  Its primary purpose is to simulate inspiraling 
and colliding black holes and neutron stars.   

For a binary black hole spacetime, $\mathtt{SpEC}$ employs a domain-decomposition, dividing the 
computational domain into about 60 elements, or ``domains".  Different domains
may have different shapes, such as cubes, cylinders, and spheres. These may in
turn have different connectivity and consequently different spectral basis-functions. 
In this paper, we only consider 
single black hole spacetimes, where the domain-decomposition consists of a set of 
concentric spherical shells.

$\mathtt{SpEC}$ employs the method of lines to evolve 
collocation point values of about 50 fundamental 
variables using a high-order Dormand Prince timestepper \cite{numericalrecipes}.  Non-linear terms 
such as those in the Einstein equations are directly computed at the 
collocation points.

Derivatives, filtering, and interpolation are performed using spectral 
transforms.  First, the collocation data is transformed to the appropriate spectral 
space. Derivatives and filtering are then implemented as operations on the spectral 
coefficients. The result is transformed back to collocation space. 

The evolved variables are tensorial, e.g. $\psi_{ab}(x^i)$.  
Here, $a,b=0,1,2,3$, indicate space-time components, and $x^i$ are the spatial 
coordinates.  Some operations, like the computation of derivatives, operate on each 
tensor-component separately. Others couple different tensor-components.  
For instance, filtering in a spherical shell relies on a representation of data
in terms of tensor spherical harmonics, to achieve a consistent truncation of 
components in angular resolution.  

$\mathtt{SpEC}$ employs the dual-frame approach \cite{scheel2006solving}.  Here, data is 
represented at collocation points at fixed grid-coordinates, the coordinate 
system in which the domain-decomposition is specified. The evolution 
equations, however, are formulated in asymptotic inertial coordinates. The two 
coordinate systems share the same time-coordinate, and their spatial coordinates 
are related by a time-dependent coordinate transformation.  

$\mathtt{SpEC}$ is a highly configurable code.  Many modules are defined 
through abstract base-classes, and are implemented in derived classes.  The 
concrete derived classes to be used for a certain domain can be chosen at 
runtime through parameter files. These choices include the coordinate mappings 
between simulation and output coordinates, which filters to implement, and  
how spectral transformations are performed (e.g. via FFTs or via a BLAS-matrix call), 
and how interpolation is performed (via a direct summation of the spectral series, 
or via a FFT onto a finer grid followed by polynomial interpolation). This 
configurability leads to many execution paths through a program.

\subsection{Programming for NVIDIA GPUs}
\label{NVIDIA}
Here we briefly introduce the pertinent characteristics of
the GPU architecture
\cite{KirkAndHwu, fermituning, keplertuning, maxwelltuning, pascaltuning}, 
as it compares with the more familiar CPU architecture.
The CPU employs a small
number ($1$-$4$ in the diagram, and up to several $10$s in contemporary 
examples) of cores that perform actual computations. All memory is
accessible by all cores. Some, much faster, memory is used to cache 
data on the grounds that repeated accesses are likely. This caching
is not managed explicitly by the user at the software level.

A relatively large amount of space is devoted to control circuitry, which for example
performs hardware-level optimizations. The extensive 
control circuity, transparent memory caching, low levels of parallelism, and
fast serial performance give the CPU great flexibility to handle a wide variety
of problems. Partly because of this flexibility, and partly because serial 
programs are much easier to optimize at the compiler level, it
is rare that a developer need think about the hardware when writing code.

The idea of the GPU is essentially to gut this entire structure and replace it 
with as many processing cores as possible. All such cores are connected to a pool of
``global" memory, which is relatively slow, though still faster than CPU 
memory. A very small global memory ``L2" cache may be present, but it is in practice
negligible compared to the CPU cache. Programmers must thus carefully manage the manner
in which global memory is accessed to achieve reasonable levels of performance, 
especially since the speed of global memory access is very often the limiting
performance factor.

The cores are organized into groups
called streaming multiprocessors (SMs). Each SM (pairs of SMs in some architectures) 
houses a small amount of control circuitry
and a hierarchy of much faster memory pools available only to it. 
These 
include another small cache, a pool of fast ``shared" memory, and a number of 
very fast registers. A thread scheduler delegates processes to individual
SMs.

Each SM has an independent control-logic, shared by all cores within each. 
This saves dye-space, but means that
all cores in an SM execute instructions in lockstep. Each core may, however,
execute these instructions upon a different global memory address. GPUs are
therefore designed to implement the SIMD (Single Instruction, Multiple Data)
model of parallelism. Programmers must take care to avoid conditional statements 
which make the instructions executed by individual cores dependent upon their 
individual data. Such statements cause the entire SM to execute each branch of the
conditional in serial.

For most of their history GPUs could be programmed only via heavily graphically-oriented
``shader" languages. 
Much more flexible access is now possible via generalized GPU frameworks
such as 
NVIDIA's 
own CUDA \cite{CUDAguide}. CUDA allows the GPU to be manipulated through a low-level C-like
interface. Parallelism is abstracted as a hierarchy of logical structures adapted
for execution by the physical structures described above. 

These logical structures
derive their names from an inconsistent and confusing metaphor with looms.
Instructions are given to the GPU by writing a CUDA \emph{kernel},
which is roughly analogous to a C function. Typically, a kernel reads an
array from global memory, performs some computation, then 
stores the results back in global memory. When executing a kernel, the programmer 
assigns a number of \emph{blocks}, and to each block a number of \emph{threads}. 
Blocks are always local to an SM and may therefore exploit SM-local resources
such as shared memory. 
Each individual thread, which has a unique index, serially executes the instructions in the kernel. The thread index can be used to address global memory,
and in this way operations on arrays can be parallelized. 

When the kernel is executed, the thread scheduler assigns the blocks to individual
SMs. The blocks are then
divided into groups of 32 threads called \emph{warps}, which execute instructions
in lockstep\footnote{In an actual loom, the term ``warp" refers to a group of 
threads which are drawn through a ``weft" of perpendicular threads held under
tension to make cloth. To our knowledge the terms ``kernel", ``block", and 
``grid" are not relevant to the textile industry.}. Each SM can simultaneously 
execute multiple warps, with the exact number depending on the specific GPU
architecture. This allows SMs to hide latency: while an instruction in a particular
warp e.g. waits for data, the SM may execute instructions from another warp
rather than simply idling.
%\begin{figure}
%\centering

  %\centerline{\includegraphics[scale=0.3, trim={0 15cm 0 0}, clip=true]{./Processor_Sketch.pdf}}
%\caption{Abstraction (image based on Figure $1$ in \cite{CUDAguide}) of CPU vs GPU 
  %architectures emphasizing differences between
%each. GPUs have vastly less control circuitry and many more processors (ALUs). 
%GPU memory is arranged hierarchically, with a large ``global" pool available
%to all processors and much slower ``shared" pools available to individual 
%``streaming multiprocessors" (the horizontal bands of cores).} 
%\label{fig:GPUarchitecture}
%\end{figure}

Writing efficient CUDA code requires low-level awareness of 
hardware. The most serious performance issue comes from the fact
that the CPU and GPU have physically different memory. 
Any data the GPU (CPU) needs from the CPU (GPU), including the kernel 
machine code itself, must be transferred over an interconnect.
Such transfers must be kept to a minimum: both in
size, since CPU-GPU interconnects are slow, and in number, since initiating a 
new transfer carries significant latency, and since all potentially-asynchronous 
GPU operations must be halted while memory is modified.

The necessity of reasoning explicitly about the low level
details of memory access 
accounts for much of GPU programming's difficult reputation. 
A warp always accesses global memory contiguously as a unit. 
Anything other than contiguous accesses in groups of 32
entries requires that the entire warp make multiple accesses. The shared memory
and registers localized within streaming multiprocessors permit comparatively
fast random access, so data can be cached here after a contiguous access to global
memory. But this must be done manually by the programmer, and the use of shared memory
in particular can easily create e.g. race-conditions. The explicit synchronizations
required to manage these, and the rather cryptic error messages supplied by 
CUDA when such management has been done improperly, can greatly complexify 
kernels.

At the kernel level the most important optimizations stem from two essential 
differences between the CPU and the GPU. First, GPUs suffer considerable
single-thread latency.  
However, these latencies can potentially be hidden by asynchronous execution. 
Ideally, then, kernels and blocks should execute for times much longer than their scheduling overhead. They should also
be numerous enough that all SMs are constantly 
occupied (but see \cite{ILP}), which also helps to hide any remaining latency.
Blocks should finally have threadsizes which
are multiples of 32 (the number of threads in a warp), since warps are indivisible
and otherwise some cores will be left idle.

Second, compared to the CPU, the GPU performs (parallel) computations far more quickly than
memory accesses. A GPU-friendly algorithm should ideally be \emph{compute-bound}, 
meaning that its \emph{arithmetic intensity}, or ratio of computations to memory accesses,
is sufficiently high that the algorithm becomes faster with increased 
computational power, rather than memory speed. In the opposite situation of
a \emph{memory-bound} kernel, speedup will be limited by the bandwidth of the 
GPU memory. This will be an important consideration when analyzing the 
performance of our implementations.

\subsection{GPU Benchmarking}
\label{Benchmarks}
Throughout this study we will often be concerned with the actual performance
achieved by GPU implementations of some algorithm compared with what
is theoretically possible. We will also be concerned with the advantage 
achieved by the GPU, relative to the CPU, when used with hardware that
is realistically available to \texttt{SpEC} users. In this subsection
we lay out our general approach to benchmarking, and thus to quantification
of such notions. 

We first estimate the potential performance of each algorithm according to
the following (rather well-established) framework. We view each algorithm
as consisting first of $M$ \emph{memory transactions}, i.e. loads and stores
of size $w$ bytes to and from RAM, resulting in $D = 10^{-9}wM$ GB of data
transacted total. All operations in this study are in double precision
and we take $w$ to be $8$ bytes throughout. 
The algorithm also performs $F$ floating-point operations
- multiplications and additions - 
a term we loosely identify with ``instructions", and which we measure in GFLOPs.
Such an algorithm can be characterized by its arithmetic intensity $I$
\begin{equation}
I \equiv \frac{F}{M}
\end{equation}
which, since GFLOPs are just numbers, is dimensionless. 

These $D$ and $F$ are what we estimate to be the \emph{minimum} possible
amounts of transacted data and floating-point operations that any
implementation of a given algorithm must perform.  We suppose
that, if run on hardware which can process data at an optimal 
bandwidth of $B$ GB/s and an optimal processing power
of $P$ GFLOP/s, an algorithm will ideally spend $D/B$ seconds on
memory transactions and $F/P$ seconds on floating-point operations.
Assuming all latency can be hidden, the total execution time would be $T=D/B+F/P$.
Considering two devices $i$ and $j$ with performance characteristics $B_i, P_i$ and $B_j, P_j$, respectively, the potential speedup from device $j$ over device $i$ is
\begin{equation}
  \label{eq:speedup}
  \mathrm{speedup} = \frac{T_i}{T_j}=\left(\frac{wB_i^{-1} + IP_i^{-1}}{wB_j^{-1} + IP_j^{-1}} \right).
\end{equation}
This also assumes that an implementation which minimizes $M$ and $F$ has
been achieved on \emph{both} devices.

When $I = I_\mathrm{eq} \equiv wP/B$, a processor will spend equal amounts
of time on memory operations and arithmetic. Put differently, an algorithmic
redesign that makes unnecessary the transmission of $D_-$ GB of data by
performing $F_+$ extra GFLOPs will be an optimization when $F_+ < I_\mathrm{eq} D_-$.
We therefore consider an algorithm to be ``memory" rather than ``compute"
bound when $I < I_\mathrm{eq}$.

Depending on whether we expect algorithms to be memory or compute bound, we
present benchmarks in terms of one of two quantities: the effective bandwidth
$BW_\mathrm{eff}$, measured in GB/s or the effective processing rate 
$P_\mathrm{eff}$, measured in GFLOP/s. Given the actual measured time $t$
required to perform the benchmarked operation, these are defined by
\begin{align}
  \label{eq:BWeffdef}
  \mathrm{BW}_\mathrm{eff} &\equiv \frac{D}{t}, \\
  P_\mathrm{eff} &\equiv \frac{F}{t}. \label{eq:Peffdef}
\end{align}
For algorithms which are heavily memory or compute bound, these quantities
will be comparable to and bound from above by $B$ and $P$ for a given device,
which allows for quick characterization of the achieved performance. More 
precisely, we can compute theoretically optimal performance metrics as
\begin{align}
  \label{eq:BWeff}
  \mathrm{BW}_\mathrm{eff, opt} = w \frac{B P}{w P + I B} \\
  P_\mathrm{eff, opt} = I \frac{B P}{w P + I B}. \label{eq:Peff}
\end{align}
Observed performance far beneath these values indicates that further
attention to optimization may be worthwhile: the algorithm as writen
may, for example, be performing extraneous operations or spending
excessive time on latency. 

\begin{figure}
  \centering
  \includegraphics[width=\textwidth]{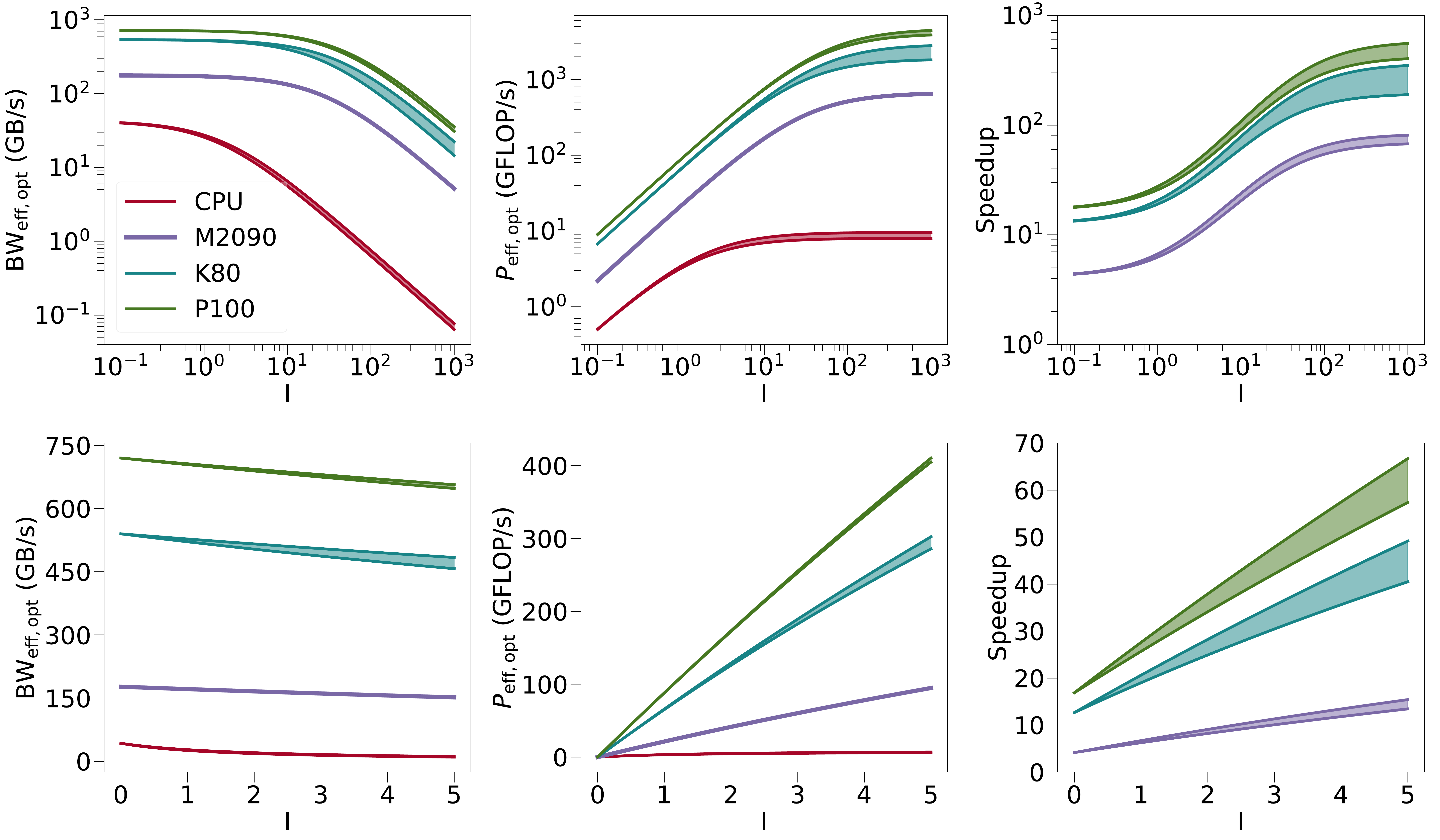}
  \caption{\textbf{Top Panels:} Effective bandwidth, processing rate, and theoretical speedup vs. a single core
    of the Intel Xeon E5-2620 as a function of arithmetic intensity in double precision for 
    the three GPUs we benchmark in this study. Improvements to arithmetic
    intensity are critical around $I=1-100$, when the speedup dependence is nearly
    linear. This assumes zero latency on both CPU and GPU, and that the 
    algorithm running on both be exactly identical; neither assumption will
    hold in practice. The ranges account for the dynamical clock 
  frequencies of all devices except the M2090. \textbf{Bottom Panels:} 
Zoom-ins around $I=0$ to $5$, relevant to the Jacobian 
contractions benchmarked later.}
  \label{fig:GPUplot}
\end{figure}

Table \ref{tab:GPUtable} shows the vendor-reported $B$ and $P$ along with an actual
measured value for $B$ obtained by running the CUDA SDK program \texttt{bandwidthTest},
which simply times the result of a device-to-device memory transfer. Using
these measured figures we also compute $I_\mathrm{eq}$ for each of four devices:
a single core of an Intel Xeon E5-2620 CPU, along with M2090, K80, and P100
GPUs. We use a \emph{single} CPU core because, as discussed earlier, \texttt{SpEC}
is incapable of \texttt{OpenMP}-style parallelization of work upon a single
domain across CPU cores. In Figure \ref{fig:GPUplot} we furthermore display
$\mathrm{BW}_\mathrm{eff}$ and $P_\mathrm{eff}$ for each processor.
We also plot the ``speedup'', i.e. the ratio between the execution time on one of the 
GPUs from Table \ref{tab:GPUtable} with that of the CPU. Larger speedups 
indicate better GPU-than-CPU performance. We will be interested
in this quantity (computed using the actually measured execution times) throughout this study.

These figures allow us to draw two immediate conclusions. First, 
the theoretical speedups range between about 4 and 400, which given
that NR runtimes are typically measured in weeks or months represents
a dramatic increase in productivity even in the pessimistic case. 
Note that actually achieved speedups may in fact be higher than
``optimal", since the CPU algorithm may not be fully optimized,
since one will typically rewrite an algorithm to achieve a 
more favourable value of $I$ during a port, and since latency
may not affect all processors equally in practice.
Second, realizing such speedups
requires high values of $I$, to a much greater or even opposite degree as would
be optimal on the CPU, as demonstrated by the approximately linear
dependence of speedup upon $I$ between $I$ of around $1$ to $100$. 
Effective porting thus often requires implementations, and sometimes
whole algorithms, to be redesigned in order to minimize memory
transactions relative to floating-point operations.

\begin{table}
\begin{center}
\begin{tabular}{ |c|c|c|c|c|c|}
  \hline
  & \multicolumn{2}{|c|}{B} & P & $I_\mathrm{eq}$ \\
  \hline
  Device & theoretical, GB/s & measured, GB/s & GFLOP/s &  \\
  \hline \hline
  CPU & 42.7 & - & 8.0 & 0.41  \\
  M2090 & 177.6 & 123 & 665.5 & 5.41 \\
  K80 (one card)& 280 & 170 & 932--1456 & 5.48--8.56 \\ 
  P100 & 720 & 449 & 4036-4670 & 9.0--10.4 \\ 
  \hline
\end{tabular}
\end{center}
  \caption{Performance specifications for our benchmarked processors. The columns
    are defined during the discussion in Section \ref{Benchmarks}. ``CPU" 
    refers to a single core of an Intel Xeon (Sandy Bridge) E5-2620. The K80 actually
    contains two separate GPUs (which share memory) on the same card. Using both requires
    similar extra effort as multi-GPU programming generally, so we profile only one throughout. 
    The K80 and P100 are also potentially
    capable of ``GPU Boost", which dynamically adjusts the core clock 
    frequency if it is possible to do so without exceeding thermal and power limits (the CPU
    has similar capabilities). 
    The ``measured" bandwidths
    were obtained by running the CUDA sample program \texttt{bandwidthTest}.
    }
    \label{tab:GPUtable}
\end{table}

\begin{figure}
\centering
\includegraphics[width=0.6\textwidth]{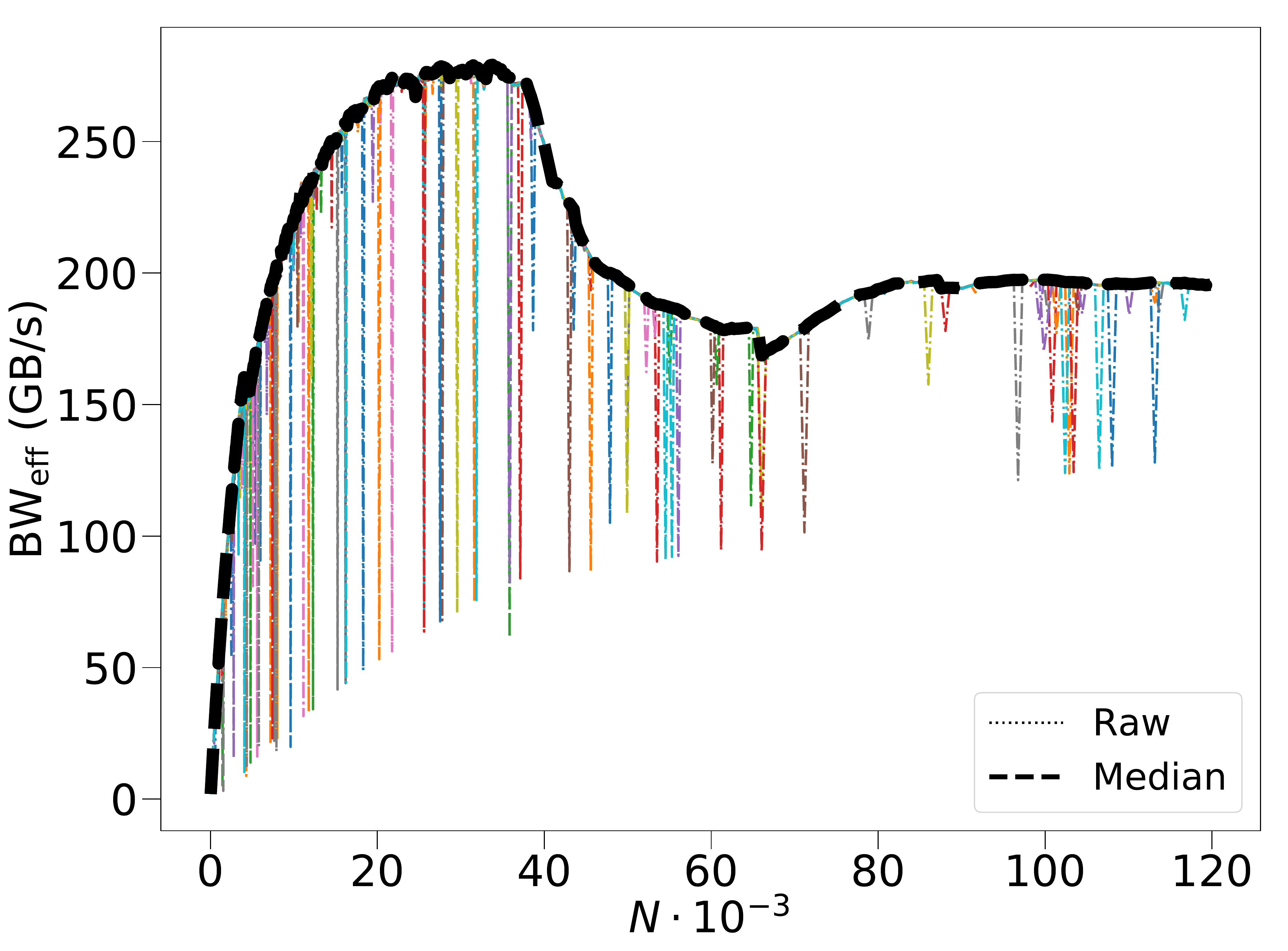}
\caption{Individual raw benchmarks from the DiffJac operation (dotted 
  coloured lines) on the P100 operating on $T_{abb}$. The median at each gridsize, 
  which clearly tracks the overall trend, is overlaid as a dashed
black line.}
\label{fig:MedianPlot}
\end{figure}

In Figure \ref{fig:MedianPlot} we, as an example, show benchmarking information
collected from the P100 GPU performing the DiffJac operation described in
Section \ref{Jacobian}. We measure performance by
$\mathrm{BW}_\mathrm{eff}$, computed
from Eq. \eqref{eq:BWeffdef}. 
This equation requires an estimate 
of the logical size of the operation $D$, which for us is 
just $8$ bytes multiplied by $M$, 
computed also in Section \ref{Jacobian}. Higher values of $\mathrm{BW}_\mathrm{eff}$
indicate better performance. We consider performance to be ``good" when
it is ``near" the estimated optimal performance $\mathrm{BW}_\mathrm{eff, opt}$
calculated from Eq. \eqref{eq:BWeff}, and plotted in Figure \ref{fig:GPUplot}.

Eq. \eqref{eq:BWeff} involves $I$ and thus both $M$ and $F$. Therefore, our 
benchmark presentations will usually take the following form. First, we will introduce
the operation to be benchmarked, describing what it does and how we have ported it.
Next, we will construct a model to estimate $M$, $F$, and $I$. Based on whether $I$
is typically less or greater than the values of $I_\mathrm{eq}$ for the GPUs in
Table \ref{tab:GPUtable}, we will present plots of either $\mathrm{BW}_\mathrm{eff}$
(when $I<I_\mathrm{eq}$, so that performance is bound most strongly by memory bandwidth)
or of $P_\mathrm{eff}$ (when $I>I_\mathrm{eq}$, so that performance is bound most
strongly by processing power). In either case, larger values indicate better 
performance. We will then discuss these plots, paying special attention
to how closely the observed performance matches the ideal performance: either 
$\mathrm{BW}_\mathrm{eff, opt}$ or $P_\mathrm{eff,  opt}$.

The machines we had access to for our K80 and P100 tests are head nodes, whose
operating systems do not employ batched processing. As a result our benchmarks
in these cases are much noisier than for the M2090 and CPU tests. 
We presume the noise to be due to machine resources being assigned to processes
besides those we seek to time. 
To correct for this
we run each test 50 separate times.  The individual benchmarks
show an obvious trend contaminated by rare, but extreme, dips in performance that
do not persist across runs. We thus take, at each gridsize, the median result of the
50 runs. To illustrate this, Figure \ref{fig:MedianPlot} plots 10 individual 
benchmarks from the DiffJac operation described in Section \ref{Jacobian}, 
with the median overlaid on top. The 
individual benchmarks mostly agree apart from isolated large spikes. The 
median tracks the agreement.

%%%%%%%%%%%%%%%%%%%%%%%%%%%%%%%%%%%%%%%%%%%%%%%%%%%%%%%%%%%%%%%%%%%%%%%%%%%%%%%
\section{Details of our port}
%%%%%%%%%%%%%%%%%%%%%%%%%%%%%%%%%%%%%%%%%%%%%%%%%%%%%%%%%%%%%%%%%%%%%%%%%%%%%%%
\subsection{Overview}
\begin{figure}
\centering
\centerline{\includegraphics[width=0.6\textwidth]{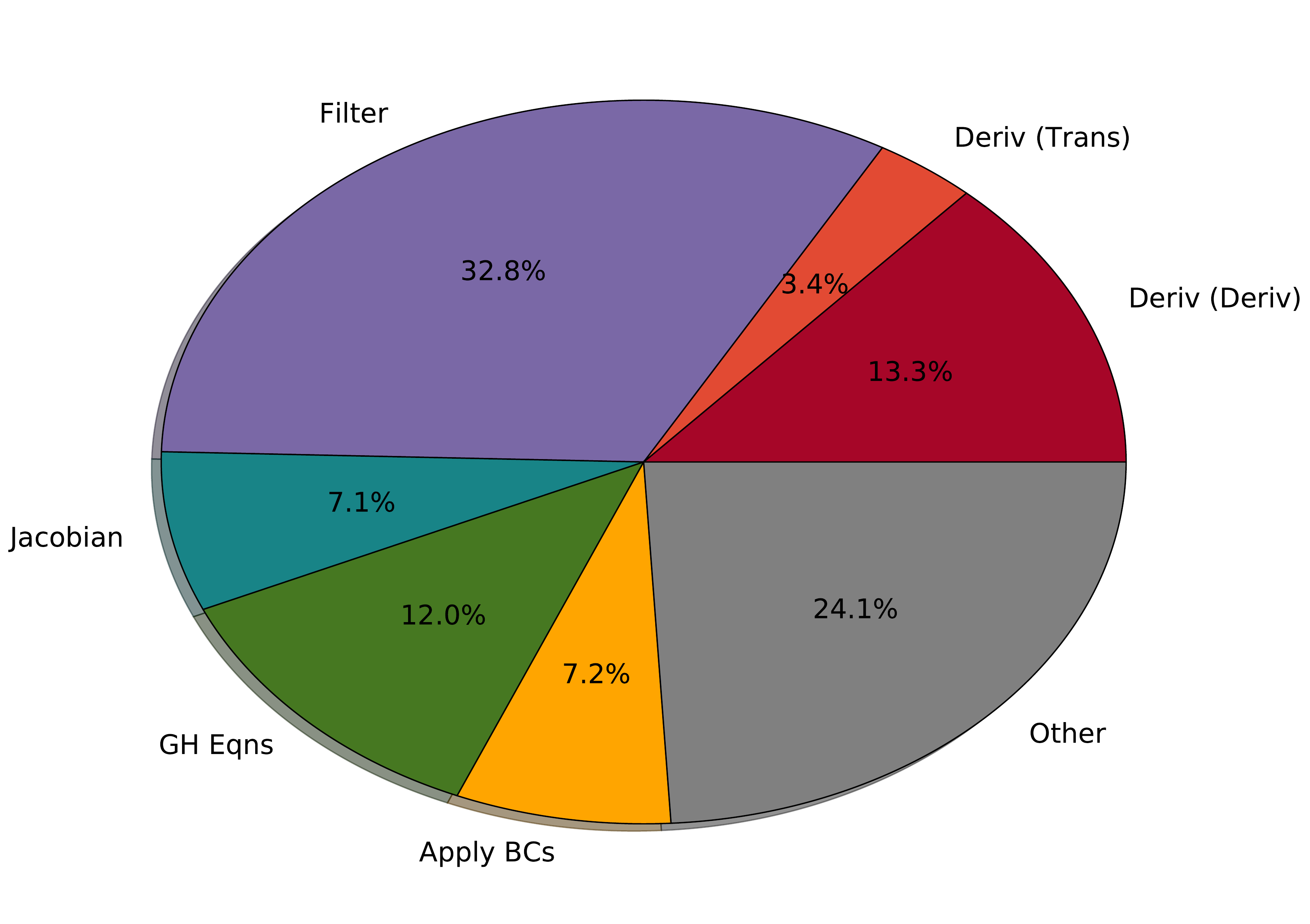}}
\caption{Illustration of the relative importance of the
modules encountered by \texttt{SpEC} during simulation of an isolated black
hole.} 
\label{fig:PieChart}
\end{figure}

We now turn our attention to \texttt{SpEC}-based black hole simulations. 
The case we consider throughout is the evolution of a single black hole.
This avoids additional complexities that occur for binaries, most
notably apparent horizon finding and MPI. This evolves 
analytically-computed Kerr-Schild initial data for an isolated
black hole, defined by its mass $m$ and its dimensionless spin-vector
$\vec{\chi}$. For a black hole with mass $m$ and angular momentum 
$\vec{J}$, one defines this dimensionless spin-vector as
\begin{equation}
\label{eq:dimensionlessspin}
\vec{\chi} = \frac{c}{G} \frac{\vec{J}}{m^2}.
\end{equation}
Black holes must have $0 \le |\vec{\chi}| \le 1$. For our test,
we choose $\vec{\chi}=(0.2, 0.3, 0.4)$, representing a moderately
spinning black hole whose spin axis is not aligned with any
of the coordinate axes. 

The spectral domains are two nested spherical shells centred on
the hole, the first extending radially from $r=1.62m$ to $r=6m$ and the second
from $r=6m$ to $r=12m$. 
This is sufficient for 
the simulation to remain stable for at least several thousand timesteps. 
A full BBH simulation would involve domains besides 
spherical shells, but spherical shells are the most important, the most individually
expensive, and the least friendly to GPU acceleration. The spherical shells in this
simulation have 10, 19, and 38 points respectively along their radial, polar, 
and azimuthal coordinates, for a total gridsize of 7220 points. 
We profile throughout from
the 5th to the 105th timestep to avoid contamination by simulation setup costs, which 
are negligible in production runs.

The pie chart of Figure \ref{fig:PieChart} illustrates, by
percentage, the per-module compute time spent by $\mathtt{SpEC}$ during such
a simulation
at a resolution comparable to that of a binary production run. These modules do the following:
\begin{enumerate}
\item Jacobian - contracts the spatial components of a tensor with a Jacobian as part of a coordinate transformation.
\item Deriv - computes the gradient of a tensor (`Deriv'). This is followed by another Jacobian multiplication (`Trans'). 
\item Filter - maps from physical to spectral space, applies a filter function, and maps back.
\item GHEqns - Computes the right-hand side of the Einstein equations.
\item Apply BCs - extracts the two-dimensional data at the boundaries of the domains, applies the boundary conditions to these, and inserts them back into the three-dimensional volume data.
\item Other - all other operations, each individually negligible.
\end{enumerate}

The rest of this section presents our approach to porting (or justifies leaving
unported) each of these.

\subsection{Memory Management}
The physically separate memory of the CPU and GPU mean that data must
be kept synchronized between the two devices: accesses to arrays in CPU 
memory must have a means to ensure that they have not been made
obsolete by a computation upon their GPU counterparts, and vice versa.
Since the actual synchronizations are very expensive, we would like
to perform them only when actually necessary. Normally this is handled 
explicitly by the user, but this approach would in our case require 
an excessive number of API calls.
We have instead 
developed C++ classes to ``lazily" hide memory management from the user. Two arrays
are maintained, one on each device, with allocations or copies made only when 
necessary. 

It turns out that GPU memory allocation is extremely expensive. 
Since \texttt{SpEC} unfortunately makes many allocations and deallocations
of memory, the naiive approach of allocating and deallocating GPU memory
in turn can massively degrade performance. 
On the other hand it is clear that repeated allocations will
be typically redundant. If an array of a given size is allocated at one timestep,
another array of the same size will very likely be allocated at the next timestep
as well. 

It therefore becomes advantageous to cache rather than deallocate GPU
pointers upon array destruction. We handle this memory cache
  with a rather straightforward map that keeps a list of pointers to
  allocated, but presently unused, GPU-memory. Initially, this cache is
  empty.  Any GPU memory allocation first checks the cache for an
  available portion of already-allocated GPU memory.  A pointer to
  such a portion, if found, 
  is removed from the cache, and returned to the
  user-code.  If no allocation is found, a new one is made using the
  usual CUDA library call.  
  
  When a portion of GPU-memory is no longer
  needed by the code, we do \emph{not} fully deallocate the memory
  via a CUDA library call. Instead, we simply add the pointer to the
  memory cache, to be reused when the next time a memory segment of
  this size is requested.  We monitor overall GPU memory usage; 
  true deallocation occurs when
the total allocated memory exceeds a certain size, or when explicitly
performed by the user.  The inspiration for this approach is
from~\cite{ModernCppDesign}, although our implementation is very
simplistic, being based on C++ standard library containers.

Usually, our array class $\mathtt{DataMesh}$ does not occur alone, but as an element 
within our container class \texttt{Tensor}. \texttt{Tensor} handles indexing in such a way
as to represent a mathematical tensor with a given dimension, rank, and 
symmetry structure. Quite often, $\mathtt{SpEC}$ will perform some operation uniformly
on all elements of a \texttt{Tensor}. Due to the kernel launch overhead GPU, 
it is much more efficient to 
handle the entire \texttt{Tensor} at once than it is to launch a new kernel for
each individual $\mathtt{DataMesh}$\footnote{For very large gridsizes, launch overhead
will be negligible compared to the runtime of individual kernels, and for
moderately large ones much of latency can be hidden by the use of concurrent
``streamed" kernels. At the gridsizes we are interested in, unfortunately, 
launch overhead remains a substantial burden.}.

$\mathtt{SpEC}$, however, allows the elements of a \texttt{Tensor} to be of arbitrary type, and
does not assume they have a uniform memory layout. This is achieved by 
handling Tensors as arrays of pointers, one per element, which 
have no relationship to one another in linear memory. On the CPU this is 
usually not a problem: calling a function once per tensor element is essentially
free, as is iterating through the \texttt{Tensor}. But on the GPU calling one kernel per
element is quite expensive. A single kernel could process the entire \texttt{Tensor},
but to do so the list of pointer-to-elements would need to be copied to the GPU,
and this copy carries again a large amount of overhead.

In practice, the capability of \texttt{Tensor}s to store nonuniform objects is rarely
used, so we handle this situation with a compromise. Each \texttt{Tensor} stores a 
list of ``\texttt{GPUPointers}", which is initially empty. When the \texttt{GPUPointers} are
explicitly asked for, they are constructed on the host and synchronized with 
the device. Subsequent accesses to the \texttt{GPUPointers} 
first check whether the \texttt{Tensor}
has been reshaped and synchronize again only if it has. This means that if the
same \texttt{Tensor} is accessed multiple times during a timestep (which often happens),
the \texttt{GPUPointers} will be synchronized only once. 

\subsection{Jacobian multiplication}
\label{Jacobian}
$\mathtt{SpEC}$ includes two performance-critical modules implementing 
coordinate changes as contractions with a 
Jacobian matrix. The first, which we call ``DiffJac", occurs 
after differentiation, bringing the derivative index into the same
coordinate frame as the tensor indices. Indices $(a, b \ldots)$ run
over spacetime and 
$(i, j \ldots)$ over space, while we represent partial
differentiation with a comma. Then 
this operation may be written symbolically as
\begin{equation}
  \label{eq:DiffJac}
  T\indices{_{ab\ldots, i}} = J\indices{_i ^j} T\indices{_{ab\ldots, j}},
\end{equation}
where $T\indices{_{ab\ldots,i}}$ is the tensor being transformed and $J\indices{_i^j}$ is the
Jacobian of the transformation. The code does not distinguish between contravariant
and covariant indices; we do so here only to clarify which indices are being summed
over. 

The second operation, which we call ``SpatialCoordJac", 
makes a coordinate change of a tensor's spatial indices
only, by contracting every possible combination of said spatial indices with
the Jacobian. First, one contraction per rank is performed over the purely
spatial indices. Next, each index is respectively set to its timelike component 
and one contraction per remaining index is performed. Subsequently, two indices are 
made timelike, followed by contractions over all unfixed indices, etc. 
In the case of a rank 2 tensor $T_{a b}$, for example,
this operation may be written
\begin{align}
  \label{eq:CoordJac}
  T_{i j} &= J\indices{_i^k} J\indices{_j^l} T_{k l}, \\ 
  \label{eq:CoordJacMid}
  T_{0 j} &= J\indices{_j^l} T_{0 l}, \\ 
   \label{eq:CoordJac2}
  T_{i 0} &= J\indices{_i^k} T_{k 0}.
\end{align}
The purely timelike component encounters a null-op, having been 
``contracted with no Jacobians". 

In practice, both of these operations are only ever applied to tensors with 
the following four rank and symmetry structures: $T_a$, $T_{ab}$, $T_{aa}$, $T_{abb}$. 
The subscripts on the above represent the rank and symmetry structure of the
tensor $T$, with repeated indices indicating a symmetry. Thus $T_{abb}$ 
indicates a dimension 4, rank 3 tensor satisfying $T_{abc} = T_{acb}$.
In the case that a symmetry structure other than one of those specified
above is encountered, our port
falls back on the CPU code.

The actual kernels are fairly simple. They could likely be optimized further,
but already perform sufficiently well that Jacobian multiplication is a very
small expense on the GPU. For the DiffJac operation of
Eq. \eqref{eq:DiffJac}, we first copy the pointer addresses of the individual tensor 
elements into linear GPU arrays. We divide the CUDA grid into two-dimension thread-blocks. The
x-coordinate runs over the spatial grid, the block index of the y-coordinate labels
the components of the input tensor, and the thread index of the y-coordinate those
of the Jacobian. Each block therefore has local to it all the Jacobian tensor pointers necessary
for the contraction, which we load into shared memory to limit
register consumption. Each thread performs the contraction in serial.

The SpatialCoordJac operation of Eqs. \eqref{eq:CoordJac}-\eqref{eq:CoordJac2} proceeds
similarly. Rather than copying individual tensor indices into a GPU array, we bundle them into a struct which
is sent to the kernel as a function argument. The pointers are then read
into device registers rather than shared memory. We thus need only one-dimensional
blocks. Register loads are much faster
than shared memory loads, and since the SpatialCoordJac operation involves 
multiple successive contractions with the same Jacobian this approach 
improves performance when the register file is sufficiently large and the
kernel is bandwidth-bound. 

\begin{figure}
\centering
\centerline{\includegraphics[width=1.1\textwidth]{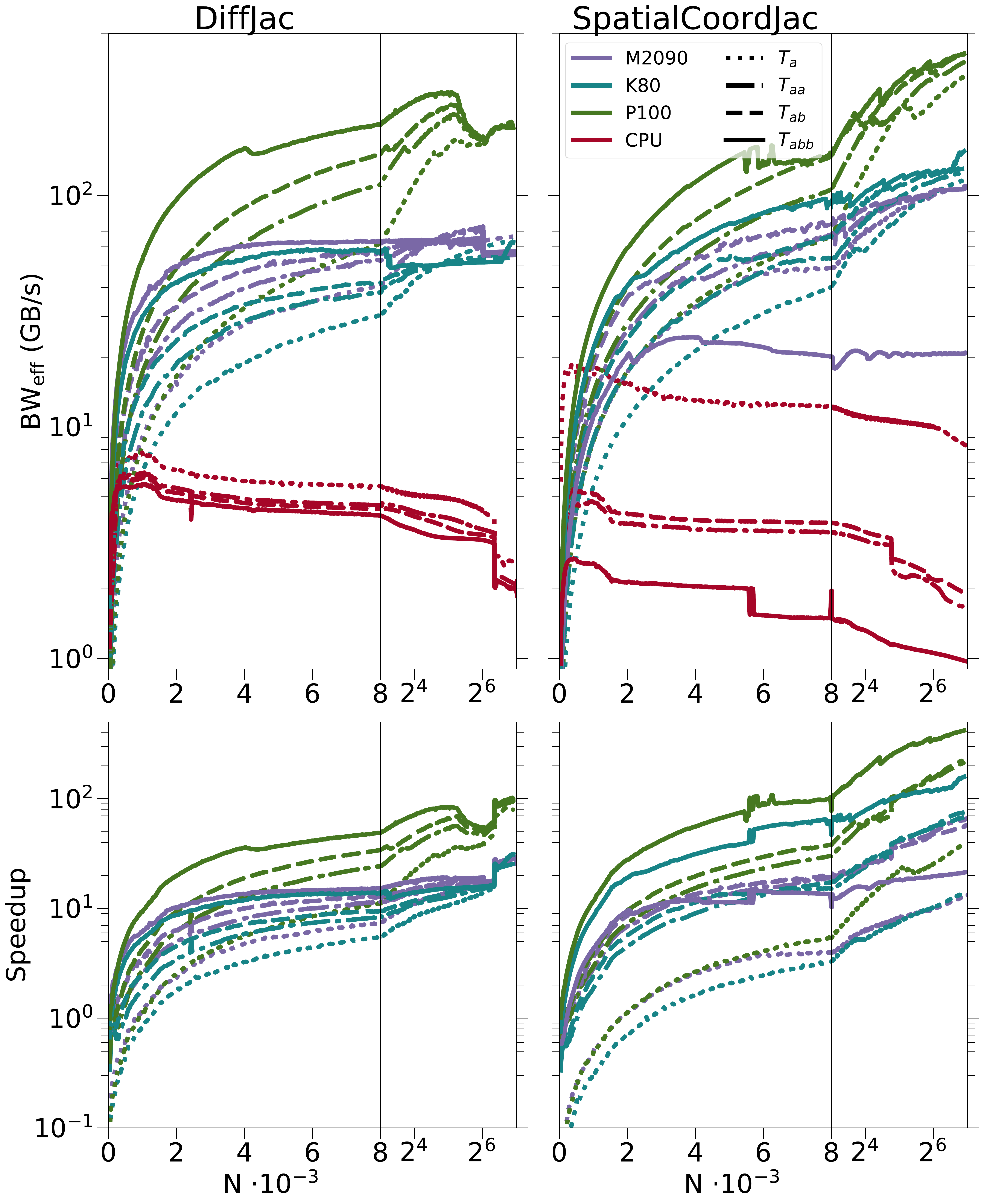}}
\caption{Effective bandwidths (top) and speedups (bottom) for the two
  Jacobian modules. Line colours and styles respectively indicate different processors
  and tensor structures. The plots switch from a linear to a logarithmic
x-axis scale at a gridsize of 8000.}
\label{fig:JacPlot} 
\end{figure}

It is nevertheless suboptimal, since the
extra register consumption can limit occupancy in practice. Most notably, for 
$T_{abb}$ on the M2090 GPU our kernel actually exhausts the available registers,
so that the local variables defined in the kernel must be allocated in global memory. 
This does not happen on the K80 and P100 GPUs, which have larger register files
(see Figure \ref{fig:JacPlot}, discussed in detail after the arithmetic intensity
models developed below, where the M2090 $T_{abb}$ 
benchmark noticeably underperforms).
In future versions we will use the shared memory approach for both operations. 
But the overall speedup is already such that this would not noticeably affect 
$\mathtt{SpEC}$'s performance (c.f. Figures \ref{fig:stackplotssbh} and \ref{fig:stackplotsbbh}).

To estimate the expected performance of these Jacobian multiplications, we need to
model their arithmetic intensity $I$, for use in Equations \eqref{eq:BWeff} and \eqref{eq:Peff}
alongside $w=8$ bytes and values of $B$ and $P$ read off from Table \ref{tab:GPUtable}.
In turn, we need to work out the number of memory operations $M$ and FLOPs $F$ 
each operation entails. The DiffJac operation 
Eq. \eqref{eq:DiffJac} must read one double per gridpoint per 
each unique array in $T\indices{_{ab\ldots, i}}$ and $J\indices{_{ij}}$, and 
then store the results again in $T\indices{_{ab\ldots, i}}$. 
We consider the arrays to consist of $N_x$ elements each, where the coordinate 
$x$ runs over physical space. Labeling the 
number of unique arrays in a tensor $T$ as $N_e^T$, and 
the spatial dimension $d$, the spatial derivative of $T$ comprises $N_e^T d$ unique
arrays, the Jacobian comprises $d^2$ unique arrays, and the minimum number of memory accesses 
is $M=N_x(2N_e^T + d) d$. The FLOP count is $F=N_xN_e^T(2d-1)d$, which can be 
computed by viewing the operation as one
multiplication per spatial gridpoint of an $(N_e^T, d)$ matrix by a $(d, d)$ matrix. 

For $d=3$ this yields an arithmetic intensity of $I=5N_e^T / (2N_e^T + 3)$ 
instructions per transaction, which notably is independent of the gridsize.
For $T_a$, $T_{aa}$, $T_{ab}$ and $T_{abb}$ respectively 
we have $N_e^T = 4, 10, 16$ and $40$, yielding  
$I_a = 1.81$, $I_{aa} = 2.17$, $I_{ab} = 2.29$, and $I_{abb} = 2.41$ 
\footnote{Here and throughout we use the subscripted notation $I_{ab...}$/$M_{ab...}$/$F_{ab...}$
to denote the arithmetic intensity/number of memory accesses/number of FLOPs of a 
particular operation working on a tensor with the subscripted symmetry structure.}
($I$ limits to $2.5$ with large $N_e^T$).
These values are only somewhat smaller than $I_\mathrm{eq}$ in Table \ref{tab:GPUtable}
for the three GPUs. We therefore expect both computational and memory throughput
to be important performance considerations. However, on the CPU $I_\mathrm{eq} = 0.41$,
indicating that computational performance is important in that case. 
Thus, we expect CPU-GPU speedups much
higher than the simple ratio between the CPU and GPU bandwidths. 
Specific ``ideal" predictions for performance and speedup are given in
Figure \ref{fig:GPUplot}. 

We now turn to the computation of $I$ for the SpatialCoordJac operation 
of Eqs. \eqref{eq:CoordJac}-\eqref{eq:CoordJac2}. In this case
$M=N_x(2N_e^T + d^2 - 2)$ memory transactions are necessary; the 
subtractive factor of 2 accounts for the purely timelike component of $T$
not participating in the operation. For $d=3$ we then have 
$M_{a} = 15N_x$, $M_{aa} = 30N_x$, $M_{ab} = 48N_x$, and $M_{abb} = 120N_x$.

The FLOP count $F$ is more complex, due to the multiple 
operations and the fact that the Jacobian is now contracted over possibly-symmetric
indices. However, we can immediately see that whatever contractions are necessary will
need to be done once per gridpoint. Therefore, $F$ will depend linearly on $N_x$, 
just as $M$ did, so that $I$ will be independent of $N_x$ for SpatialCoordJac,
just as it was for DiffJac. 

We nevertheless seek concrete estimates of $F$ for each tensor structure.
If symmetries can be neglected, $F$ for a tensor
of arbitrary rank $r$ can be computed without much difficulty. 
The initial spatial contraction, Eq. \eqref{eq:CoordJac}, can be
viewed as $r$ matrix-multiplications between the $(d, d)$ Jacobian and the 
$(d, d^{r-1})$ input tensor. This takes $N_x r d^r (2d-1)$ operations. One index 
is then (Eq. \eqref{eq:CoordJacMid}) made timelike, followed by another set of contractions upon an input 
tensor of rank $r-1$. There are $r$ different ways to set one index timelike,
so this second step is done $r$ times. Eq. \eqref{eq:CoordJacMid} 
therefore takes $N_x r(r-1)d^{r-1}(2d-1)$
operations. Following through this reasoning for the full series of operations,
suppose $W(r, j)$ is the number of unique arrangements of $j$ zeros in a tensor of rank $r$, 
and $S(r, j)$ is the
number of unique spatial components in the tensor so fixed. 
Then the operation count (for any tensor) is
\begin{equation}
  \label{eq:CoordJacGen}
  F = N_x (2d-1) \sum_{j=0}^{j=r-1}W(r, j) S(r, j) (r-j),
\end{equation}
which, for a tensor with no symmetries, reduces to
\begin{equation}
  \label{eq:CoordJacNoSym}
  F = N_x (2d-1)\sum_{j=0}^{j=r-1}\binom{r}{j} d^{r-j}(r-j).
\end{equation}
This is a rather steep function of $r$. In $d=3$, Eq. \eqref{eq:CoordJacNoSym} 
it gives $F_{a} = 15N_x$ and $F_{ab} = 120N_x$. 

The combinatorics when symmetric index pairs are allowed are much more involved. 
Fortunately we are only interested in the simplest two such quantities, $F_{aa}$ and 
$F_{abb}$. 
A tensor of rank $r$ with $\sigma$ symmetric index pairs has $S(r, 0) = d^{r-2\sigma} \binom{d+1}{2}^\sigma$. The symmetric pair also reduces the number of unique ways to fix indices. For $r=2, \sigma=1$ we have 1 term with $S = \binom{d+1}{2}$ and 1 term with $S=d$. In total this gives $F_{aa} = N_x \left(\binom{d+1}{2} 2 (2d-1) + d (2d-1)\right) = 75N_x$. 

Finally, for $r=3, \sigma=1$, we get one purely spatial arrangement with $S(r, 0) = d\binom{d+1}{2}$. There are two ways to fix only one index. Fixing the non-symmetric index gives a contribution equal to that for $T_{ab}$, while fixing part of the symmetric pair gives the same from $T_{aa}$. Either of the two ways of fixing two indices gives a $T_a$ contribution. In total, we have $F_{abb} = N_x d\binom{d+1}{2}3(2d-1) + F_{ab} + F_{aa} + 2F_{a} = 495N_x$. 

The arithmetic intensities for SpatialCoordJac are thus 
$I_a = 1$, $I_{aa} = 2.5$, $I_{ab} = 2.5$, and 
$I_{abb} = 4.125$ ($I$ in general depends on the symmetry structure). 
Since these numbers are very similar to the ones we obtained for DiffJac,
we expect similar performance in both cases. In particular, performance
will depend most strongly upon hardware memory bandwidth $B$ on the GPU, and
on processing power $P$ on the CPU, and $\mathrm{BW}_\mathrm{eff}$ will be
an appropriate metric of performance. 

With these theoretical considerations in mind, we now turn to
actual benchmarks. On each of the four architectures listed in
Table \ref{tab:GPUtable}, we execute DiffJac and SpatialCoordJac upon
tensors of structures $T_a$, $T_{aa}$, $T_{ab}$, and $T_{abb}$.
Using the models above and the measured execution time $t$, 
we then compute $\mathrm{BW}_\mathrm{eff}$ for each case from 
Eq. \ref{eq:BWeffdef} 
\footnote{As described in Section \ref{Benchmarks} and illustrated in
Figure \ref{fig:MedianPlot}, we clean the $\mathrm{BW}_\mathrm{eff}$
measurements across multiple executions by taking their median.}.
These $\mathrm{BW}_\mathrm{eff}$ results are plotted against the
spatial gridsize $N_x$ in the top
panels of Figure \ref{fig:JacPlot}, whose $x$ axis switches from linear
to logarithmic at $N_x=8000$ in order to compactly display the large-$N_x$
behaviour. Each line on those plots represents a different processor
from Table \ref{tab:GPUtable}, indicated by differing colours, or
a different tensor structure, indicated by differing linestyles.
The bottom panels show the CPU-GPU speedup, i.e. the ratio between
the execution time on the indicated GPU and that on the CPU. For
both $\mathrm{BW}_\mathrm{eff}$ and the speedups, higher $y$-axis
numbers indicate superior GPU performance.

We now highlight the salient features of Figure \ref{fig:JacPlot}
and interpret them in light of our theoretical expectation from
the computation of $I$ and the pragmatics of our implementation.
Our predicted arithmetic intensities $I$ were between $1$ and $5$.
Consulting the bottom-left panel of Figure \ref{fig:GPUplot}, we
see that at peak performance $\mathrm{BW}_\mathrm{eff}$ should thus be 
roughly constant in $I$, and thus independent of both the tensor structure
and of whether we are performing DiffJac or SpatialCoordJac. 

The expected independence of tensor structure and $N_x$ can, for the GPUs, 
be seen in Figure \ref{fig:JacPlot}.
There, the various linestyles representing differing tensor structures
appear for each GPU to converge at large $N_x$. The 
exception of the $T_{abb}$ kernel running SpatialCoordJac on the M2090,
whose performance can be seen from the top right panel of Figure \ref{fig:JacPlot}
to be far beneath the other M2090 curves. In this case, the kernel spills
registers into local memory. The CPU curves show a much stronger dependence
upon tensor structure than predicted when running SpatialCoordJac, and in both
cases show a clear negative dependence upon $N_x$, presumably since large gridsizes
overflow the CPU cache.

For both DiffJac and SpatialCoordJac, the P100 outperforms the M2090 by
about a factor of $3$ at large $N_x$, as can be seen by comparing the
rightmost edges of the $\mathrm{BW}_\mathrm{eff}$ lines in Figure \ref{fig:JacPlot}
for these processors. This is consistent with the expectation
from a similar comparison from the bottom left panel of Figure \ref{fig:GPUplot}.
These performances are, however, quantitatively each about a further factor K80
of 3 away from Figure \ref{fig:GPUplot}'s prediction, indicating that 
algorithmic redesign could likely further improve performance. This is especially
true for the K80, which actually performs slightly less well than the M2090
despite greatly superior specifications. However, as will be shown in
Section \ref{OverallBenchmarks} (specifically Figures \ref{fig:stackplotssbh} and 
\ref{fig:stackplotsbbh}), the speedups we have already obtained make
Jacobian multiplications a very small part of the overall black hole
simulation runtime, so further effort would have little practical
impact.

The GPU performance for SpatialCoordJac at large gridsize is, 
contrary to the expectation from the above analysis (i.e. from the 
fact that $I$ is quite similar for both operations), 
consistently about a factor of 2-5 better than for DiffJac,
as can be seen by comparing the large $N_x$ behaviour of identically
styled curves for the GPUs on the left and the right panels of Figure
\ref{fig:JacPlot}. 
The relevant kernels are coded somewhat differently: 
all the necessary pointers-to-tensor elements are 
first collected into a struct, which is passed to the 
GPU as an argument to the kernel rather than by a CUDA memory copy. 
All necessary data are then loaded into registers in a way that
interleaves memory accesses with computations. 
This approach may ultimately entail fewer, 
or better optimized, memory accesses, since no explicit pointer 
indirections are coded in. The 
staggered instructions may also improve latency, since less time 
need be spent waiting for data.

\subsection{Spectral Operations: Differentiation and Filtering}
\label{spectral}
$\mathtt{SpEC}$ is named for its use of the pseudospectral collocation method 
\cite{boyd2013chebyshev, fornberg1998practical, canuto1988spectral, canuto2007spectral, Bonazzola:1998ge}. In this section we describe our porting strategy for two operations,
differentiation and filtering, which make explicit use of these methods. Let us
begin by introducing spectral methods in the simplest case of $1\mathrm{D}$ PDEs and
scalar variables. Spectral methods represent
the solution $u(x)$ as a series expansion in basis functions $T_k(x)$:
\begin{equation}
  \label{eq:SpectralExpansion}
  u(x) = \sum_{k=0}^{N-1} \tilde{u}_k T_k(x).
\end{equation}
The $\tilde{u}_k$ above are called spectral coefficients. The approximation
arises because $N$ is finite. 

Furthermore, there is a set of collocation points
\begin{equation}
  \label{eq:collocation}
  x_i, i=0,\ldots,N-1.
\end{equation}
The function values at the collocation points, $u_i \equiv u(x_i)$, can
be computed by a matrix multiplication:
\begin{equation}
  \label{eq:SpecToPhys}
  u_i = \tilde{M}\indices{_i^k} \tilde{u}_k,
\end{equation}
where $\tilde{M}\indices{_i^k} = T_k(x_i)$. For suitable choice of 
collocation points, the inverse is also a linear transformation:
\begin{equation}
  \label{eq:PhysToSpec}
  \tilde{u}_k = M\indices{_k^i}u_i.
\end{equation}
We shall refer to Eq. \eqref{eq:SpecToPhys} as ``SpecToPhys"
and to Eq. \eqref{eq:PhysToSpec} as ``PhysToSpec".

For a Fourier series and Chebyshev polynomials, the transforms Eq. \eqref{eq:SpecToPhys}
and Eq. \eqref{eq:PhysToSpec} can be evaluated respectively with a fast-Fourier-transform
or fast-cosine-transform (we hereafter loosely use the term
``FFT" to refer to both possibilities), with $\mathcal{O}(N \log{N})$ complexity 
scaling rather than the naiively-expected $\mathcal{O}(N^2)$ for matrix-vector
multiplication.

The advantage of spectral methods is that many operations of interest, most
notably including differentiation, can also be performed with an FFT,
yielding $\mathcal{O}(N \log{N})$ complexity scaling overall. Since,
for example, the basis functions form a complete set, their derivatives
are linear combinations of basis functions
\begin{equation}
  \label{eq:Deriv}
  T'_k(x) = \sum_k^{N-1} \tilde{D}\indices{_k^l} T_l(x)
\end{equation}
which we can exploit to find the differentiation matrix $\tilde{D}\indices{_k^l}$
analytically. Multiplying this matrix by the spectral coefficients 
$\tilde{u}_k$ - which, again for a Fourier series and Chebyshev polyomials,
can be done with a $\mathcal{O}(N \log{N})$ FFT - yields $\tilde{u}'_k$, the 
spectral coefficients of the solution's derivative. The real space derivative $u'_i$
can then be obtained using Eq. \eqref{eq:SpecToPhys}, with FFT-like scaling overall. 

Differentiation accounts for around 20\% of $\mathtt{SpEC}$'s total runtime during the 
SingleBH test (c.f. Figure \ref{fig:PieChart}). A second operation called the 
spectral filter consumes about an additional 30\%, and is very similar in form. Here,
we perform the same spectral transformations as in \eqref{eq:PhysToSpec} and
\eqref{eq:SpecToPhys}, but the matrix in \eqref{eq:Deriv} is designed to apply 
some filtering transformation to the spectral coefficients rather than
to compute derivatives. For example, the Heaviside filter zeros out all Fourier
modes with a frequency above a certain value.

In practice, $\mathtt{SpEC}$ maintains for each one of these steps a complicated battery
of C++ classes that are appropriate for different choices of simulation domain,
spectral basis function, and low-level implementation details. 
Producing hand-written CUDA equivalents of all possible execution pathways would take, to
say the least, significant effort. In particular code maintenance would become
unmanageable. To avoid this we capitalize on three features of the spectral operations.
First, they are all linear maps. Second, a given process will encounter only
manageably few unique instances of each. Third, in a practical simulation 
each unique instance will be encountered by a process very many times.

These properties in concert make feasible a general strategy based on tracing
out an explicit matrix representation for each function. Specifically,
considering for example the differentiator, we express the entire transformation
as an explicit multiplication with a \emph{single} matrix 
$D\indices{_j^i} = (\tilde{M}DM)\indices{_j^i}$.
We then have

\begin{equation}
  \label{eq:Differentiation}
  u'_j = D\indices{_j^i} u_i
\end{equation}
which is mathematically
equivalent to the differentiation operation, however the latter is implemented. 
We can exploit
this fact to trace out an explicit matrix representation of $D\indices{_j^i}$.
Specifically, we set $u_i = \delta_{ik}$ for some $k$, 
where $\delta_{ik}$ is the Kronecker
delta, and then pass this input through the actual extant CPU code. The result
is the $k$th column of the matrix $D\indices{_j^i}$. By repeating this procedure
for all $k$'s, we trace out the entire matrix in $N_x$ function calls. 

Having traced out $D\indices{_j^i}$, we maintain an associative array (which we
call a dictionary) between 
it and whatever function input specifies a mathematically-new transform.
For the differentiator, this is the gridsize and derivative index, while 
for the spectral filter, it is the gridsize, the particular
filter function to be applied, and in some cases the tensor structure of the input.
The extra function calls needed to build the matrix are in practice very few 
compared to the 
full number that will be made over the $\mathtt{SpEC}$ runtime, so the extra expense
can be ignored. 

In general, implementing the spectral operations by explicit matrix multiplications
such as Eq. $\eqref{eq:Differentiation}$ will result in worse asymptotic complexity
than is achieved by the spectral CPU code, whose expense is dominated by either an FFT 
or a closely related algorithm. Nevertheless, this approach can be advantageous, especially
when viewed as a CPU-to-GPU porting strategy. First, instead of needing to port, optimize,
and maintain a parallel GPU code for each of the very many possible
spectral transformations, only one or a small number are necessary. Second, the operation
counts at low-$N$ can be such that matrix multiplications actually outperform
fast transforms at practical gridsizes. Third, FFTs are, in general, much more difficult
to parallelize than matrix multiplications, leading to much lower FLOP/s 
for the former: for example NVIDIA reports large-$N$ double precision operation 
rates of around 150 GFLOP/s for their cuFFT library running on a K40 GPU \cite{cuFFT}, 
compared to near-peak performance
in the TFLOP/s regime for matrix multiplication \cite{cublas} (of course, the FFT involves many
fewer operations). Finally, matrix multiplication can be performed by the (cu)BLAS
function \texttt{dgemm}, which is possibly the most heavily optimized function in existence. 

Let us now consider the realistic case of 3D grids and tensorial solution variables.
Usually, the independent tensor components are decoupled,
and so generalizing to tensors with $N_e$ independent components simply 
involves a factor of $N_e$ extra function calls. But the higher
spatial dimensions are qualitatively important, 
since they change the shapes and characters
of the matrix multiplications (or fast transforms).

Let us now consider our port of the differentiator as it works in practice.
We start with a function $u(x, y, z)$ available at physical collocation points
$u_{ijk} = u(x_i, x_j, x_k)$, and denote by $N_x$, $N_y$, and $N_z$ the physical
gridsizes in the subscripted dimension \footnote{While our discussion centres upon
$d=3$, generalization to other dimensions will be obvious.}. The full domain topology is an outer product of 
so-called ``irreducible topologies" which cannot be themselves expressed 
as outer products, and each irreducible topology 
will be associated with its own set of spectral basis functions. These
basis functions will depend on each physical coordinate on their associated
domain. For example, on an $\mathcal{I}1\otimes \mathcal{I}1 \otimes \mathcal{I}1$ domain,
where the irreducible topology $\mathcal{I}1$ is that of a closed line segment,
we have three sets of spectral basis functions, and each set depends on 
only one physical coordinate; we thus call the basis functions 1D. In this case
we write
\begin{equation}
  u(x, y, z) = \sum_{i,j,k}^{N_x, N_y, N_z} \tilde{u}_{ijk} T_i(x) T_j(y) T_k(z).
\end{equation}
In particular, we can obtain e.g. the $\tilde{u}_i$ coefficients, which are all
we need to compute the $x$-derivative, without performing the other two sums:

\begin{equation}
  \tilde{u}_{\alpha k j} = M\indices{_\alpha^i} u_{ikj}.
\end{equation}
The derivatives are again linear combinations of basis functions, we again
finish by mapping back to the collocation points, and we again can trace out
an explicit matrix representation of the entire operation by feeding delta
function input through the CPU code. Denoting this matrix representation by the
capital letter corresponding to the physical coordinate upon which it operates, 
we have

\begin{equation}
  u_{abc,d} = \delta_{d1}X\indices{_a^i} u_{ibc} 
  + \delta_{d2} Y\indices{_b^j} u_{ajc}  
  + \delta_{d3} Z\indices{_c^k} u_{abk}. 
\end{equation}

In some cases \texttt{SpEC} works upon domains composed of irreducible topologies
which are not 1D in the above sense - that is, their associated spectral 
basis functions depend on more than one physical coordinate. The most notable
example is spherical shells, with topology $\mathcal{I}1 \otimes \mathcal{S}2$.
The $\mathcal{I}1$ irreducible topology, representing the radial direction $r$,
admits a spectral basis of 1D Chebyshev polynomials that depend only on $r$,
but the spherical harmonics $Y_{lm}$ depend on both angular coordinates,
$\theta$ \emph{and} $\phi$. 
\texttt{SpEC} furthermore in this case uses a compressed, but slightly redundant, 
spectral 
representation, so that $\tilde{M}\indices{_i^k}$ is not simply the inverse of 
$M\indices{_k^i}$. The matrix product $\tilde{M}\indices{_i^k} M\indices{_k^j}$
projects into a subspace of $\{u_i\}$. 

In this case, we have physical variables $u(r, \theta, \phi)$ available
at physical collocation points $u_{klm} = u(r_k, \theta_l, \phi_m)$.
The physical gridsizes in the $r$, $\theta$, and $\phi$ directions will
be denoted $N_r$, $N_\theta$, and $N_\phi$, while the number of $l$ 
modes
maintained in the spectral representation
will be called $N_l$ (the number of Chebyshev coefficients is just $N_r$).
The spectral coefficients are written
\begin{equation}
  u(r, \theta, \phi) = \sum_{k=0}^{N_r}\sum_{l=0}^{N_l}\sum_{m=-l}^{l} \tilde{u}_{klm}\; T_k(r)\,Y_{lm}(\theta, \phi),
\end{equation}
where $Y_{lm}(\theta,\phi)$ represents the spherical harmonics, and $T_k(r)$ a Chebyshev polynomial operating on a suitably rescaled radial coordinate $r$.  Because $Y_{lm}(\theta,\phi)$ depend on both the $l$ and $m$ index, computation of either angular derivative requires the entire double sum
over both $l$ and $m$. We end up with

\begin{equation}
  \label{eq:DifferentiatorI1S2}
  u_{abc,d} = \delta_{d1}R\indices{_a^i} u_{ibc} 
  + \delta_{d2} \Theta\indices{_{bc}^{jk}} u_{ajk}  
  + \delta_{d3} \Phi\indices{_{bc}^{jk}} u_{ajk}. 
\end{equation}
There is now the practical business of expressing these operations as sequences
of BLAS calls \footnote{BLAS accepts `transpose' parameters which determine whether the 
  input matrices are to be read in standard ('N') or transposed ('T') format. $\mathtt{SpEC}$
  stores physical data in row-major format, but BLAS assumes column-major, so
  the input is implicitly transposed anyway. By chance, this naturally leads
  to the choice 'N','N' for the first basis function and 'T','N' otherwise,
which are the two most favourable cases in terms of performance.}
. $\mathtt{SpEC}$ stores the collocation data $u_{ijk}$ as physically contiguous
arrays, so that we are free to join together adjacent indices. We may thus view
the data equivalently as a matrix $u_{i, j:k}$ (for the first transform), 
as a matrix $u_{i:j, k}$ (for the last, or the last two in the $\mathcal{I}1 \otimes \mathcal{S}2$ case), or 
as a set of $k$ submatrices $u_{i, j}$ (for the middle transform), each one of which
is multiplied by the appropriate transformation matrix. The colon notation above
indicates vectorization into a single index; i.e. $u_{x_i, y_j:z_k}$ is a matrix
of size $(N_x, N_y N_z)$.

Since these are just sequences of matrix multiplies, we can easily estimate
FLOP and memory transaction counts for each. For $\mathcal{I}1\otimes \mathcal{I}1 \otimes \mathcal{I}1$,
we first shape the input
as an $(N_x, N_y N_z)$ matrix and multiply with the $(N_x, N_x)$ x-transform 
matrix.
Next we  shape the input into $N_z$ $(N_x, N_y)$ submatrices and multiply
each with the $(N_y, N_y)$ y-transform matrix. Finally, we shape it into an
$(N_x N_y, N_z)$ matrix and multiply that with the $(N_z, N_z)$ z-transform
matrix. These ``reshapings" are just parameter choices to 
BLAS, and involve actual copies. We repeat this procedure once
for each of the $N_e$ independent components of the input tensor.
The operation count for any particular coordinate $x_d$ with size $N_d$ 
has the functional form
\begin{equation}
  \label{eq:I1Fx}
  F = N_e N_x N_y N_z (2 N_i - 1). 
\end{equation}
The x and z transforms involve

\begin{equation}
  \label{eq:I1Mx}
  M = N_e(N_d^2 + 2N_xN_yN_z)
\end{equation}
memory operations. As formulated above, however, the y-transform matrix must be
read in by the device $N_z$ times, with each read acting upon a fraction $1/N_z$ 
of the entire volume data. We thus have
\begin{equation}
  M = N_eN_yN_z(N_y + 2N_x).
\end{equation}
These memory access estimates somewhat exceed what is 
strictly required. The transform matrices, for example, are the same for
each tensor component, and a kernel could load them from global
memory only once for the entire tensor. Optimizing such a kernel to 
outperform cuBLAS even given the extra accesses would, however, be 
difficult, especially since matrix multiplication is compute-bound. 
If $\mathtt{SpEC}$ stored entire tensors as 
contiguous arrays this could be achieved 
using the cuBLAS function
\texttt{cublasdgemmStridedBatched}. Since this is not in fact
the case, we use the above accounting.

In total, we have
  \begin{align}
    \label{eq:I1total}
    M &= N_e(6N_xN_yN_z + N_x^2 + N_z N_y^2 + N_z^2) =  N_e(7N^3 + 2N^2) \\ 
    F &= N_eN_x N_y N_z [ 2(N_x + N_y + N_z) - 3] = N_e(6N^4 - 3N^3).
  \end{align}
where the rightmost equalities assume $N_x=N_y=N_z \equiv N$, in which case the 
arithmetic intensity is $I = N \frac{3(2N-1)}{7N+1} \sim \frac{2}{7} N$ 
(note again that $N$ here is the \emph{linear} gridsize). 
Comparing with $I_\mathrm{eq}$ from Table \ref{tab:GPUtable}, we see this 
operation will be compute-bound at any realistic input size.  
For $\mathcal{I}1\otimes\mathcal{S}2$, the transform matrix shapes 
are $(N_r, N_r)$ for the radial
transform and $(N_\theta N_\phi, N_\theta N_\phi)$ for both of the angular ones.
The input reshapings are $(N_r, N_\theta N_\phi)$ and $(N_\theta N_\phi, N_r)$. 
Due to the dependence of the spherical harmonic basis functions on \emph{both} 
angular coordinates,
the transforms are also over both, even if we only seek e.g. the
$N_\theta$ derivative. For the same reason,
we do not break into $N_\phi$ submatrices for the $N_\theta$ transform, as we
did for $N_y$ in $\mathcal{I}1\otimes\mathcal{I}1\otimes\mathcal{I}1$. 

\begin{figure}
\centering
\centerline{\includegraphics[width=0.8\textwidth]{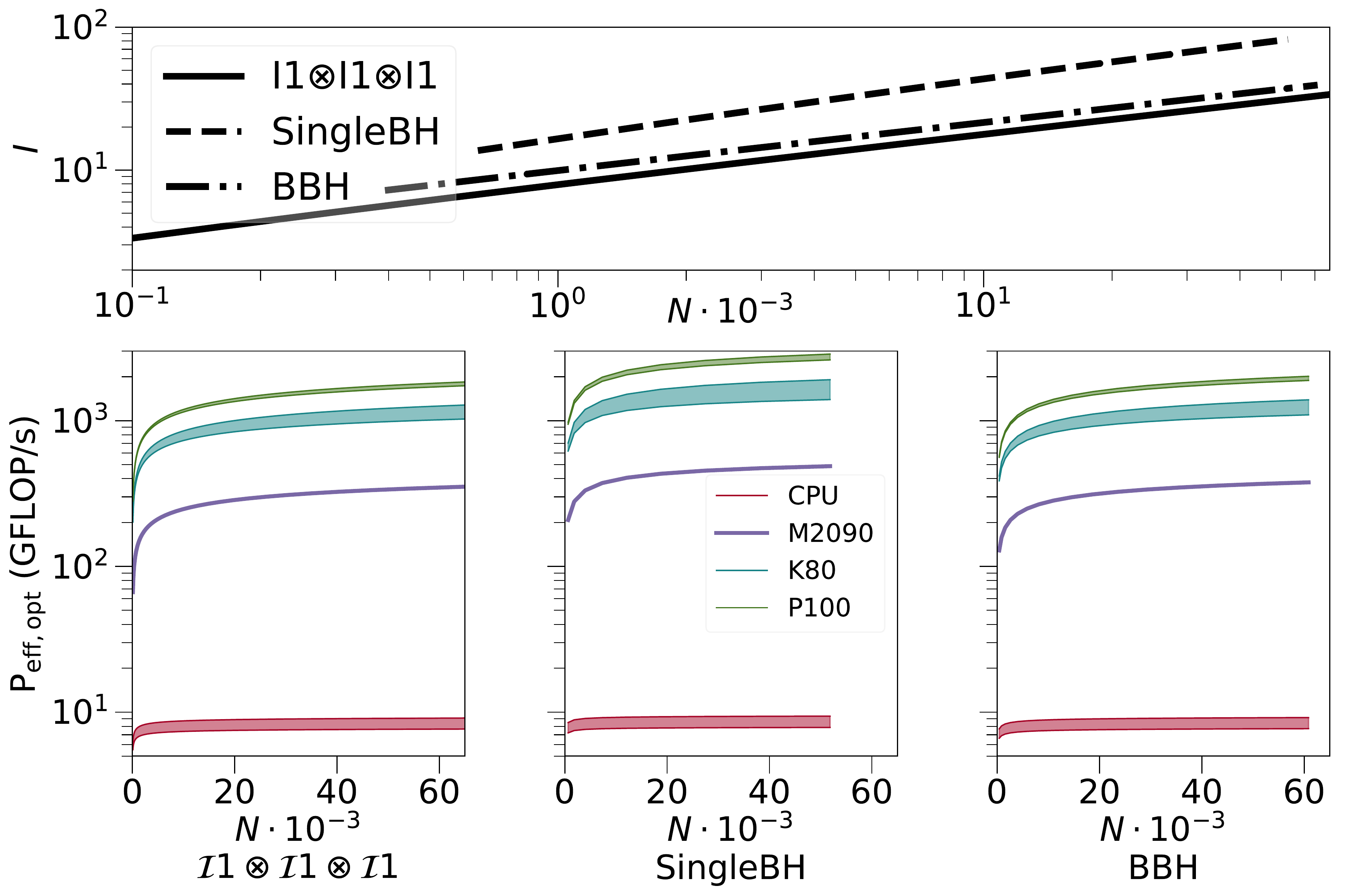}}
\caption{Top: Arithmetic intensity $I$ plotted as a function of gridsize $N$ for
the matrix multiply differentiator operating on 
$\mathcal{I}1 \otimes \mathcal{I}1 \otimes \mathcal{I}1$, and upon $\mathcal{I}1 \otimes \mathcal{S}2$
with the SingleBH and BBH gridpoint distributions. For 
$\mathcal{I}1 \otimes \mathcal{I}1 \otimes \mathcal{I}1$ these estimates also 
apply to the spectral filter.
Bottom: theoretical zero-latency processing rate $P_{\mathrm{eff, opt}}$ for each of the 
benchmarked devices as a function of gridsize.
Note that the CPU uses a more efficient (at large gridsize) algorithm, 
so these lines do not bound its performance.}
\label{fig:difftheory}
\end{figure}

Noting that $N_i = N_\theta N_\phi$ for the angular transforms, $F$ and $M$
have the same forms as in \eqref{eq:I1Fx} and \eqref{eq:I1Mx}. In total, 
we have
  \begin{align}
    M &= N_e(N_r^2 + 2N_\theta^2 N_\phi^2 + 6N_r N_\theta N_\phi) \\ 
    F &= N_e N_r N_\theta N_\phi[2N_r + 4N_\theta N_\phi - 3].
  \end{align}
In practical simulations, the resolutions $N_r$, $N_\theta$, and $N_\phi$ can 
differ widely from one another. For our
benchmarks we therefore distribute points by two prescriptions.
The ``SingleBH" benchmarks are on spherical shells that mirror those found in
$\mathtt{SpEC}$'s isolated black hole evolutions. Resolution is 
controlled by a resolution parameter
$k=0,1,\ldots,10$, in terms of which we have $N_r = 9+4k$, $N_\theta = 6 + 2k$, $N_\phi = 4k+12$,
and thus $N_\theta N_\phi = 8k^2 + 48k + 72$.  The ``BBH" benchmarks use roughly
the same point distribution
as used initially for the spherical shells closest to the apparent horizons
of black hole binaries in an actual BBH simulation \footnote{$\mathtt{SpEC}$ employs adaptive
  mesh refinement during a run, so the actual point distribution during a 
simulation may be rather different.}. The radial resolution is comparatively
much lower in this case. Specifically we have $N_r = 4+k$, $N_\theta = 7 + 2k$, 
$N_\phi = 4k + 14$, $k=1,2,\ldots,16$, and thus
$N_\theta N_\phi = 8k^2 + 56k + 98$. For SingleBH, we have

  \begin{align}
    M &= N_e(128k^4 + 1728k^3 + 8512k^2 + 18216k + 14337) \sim 128N_ek^4\\ 
    F &= N_e(1024k^5 + 14848 k^4 + 85536k^3 + 244728k^2 + 347760k + 196344) \sim 1024N_e k^5.
  \end{align}
The arithmetic intensity $I=13.7$ for $k=0$
and grows approximately linearly thereafter. For the BBH benchmarks we have 
  \begin{align}
    M &= N_e(128k^4 + 1840k^3 + 9937k^2 + 23892k + 21576) \sim 128 N_e k^4 \\ 
    F &= N_e(256k^5 + 4624k^4 + 33368k^3 + 120252k^2 + 216426k + 155624) \sim 256N_ek^5.
  \end{align}
This arithmetic intensity $I$ starts at $7.2$. The operations will clearly be 
compute-bound in all cases. Of course, equal $k$ implies different total 
gridsize between cases, so it is difficult to estimate performance
at equal gridsize from the above. To do that we refer to Figure 
$\ref{fig:difftheory}$, where $M$, $F$, and $I$ are shown
for each of the three grids. With $I$ in hand as a function of gridsize, 
we can refer once more to 
Figure \ref{fig:GPUplot} to predict ideal effective processing rates
$P_{\mathrm{eff}, \mathrm{opt}}$ from Eq. \eqref{eq:Peff}, which we plot
against gridsize in the bottom panels of Figure \ref{fig:difftheory}. 

\begin{figure}[t]
\centering
\centerline{\includegraphics[width=0.9\textwidth]{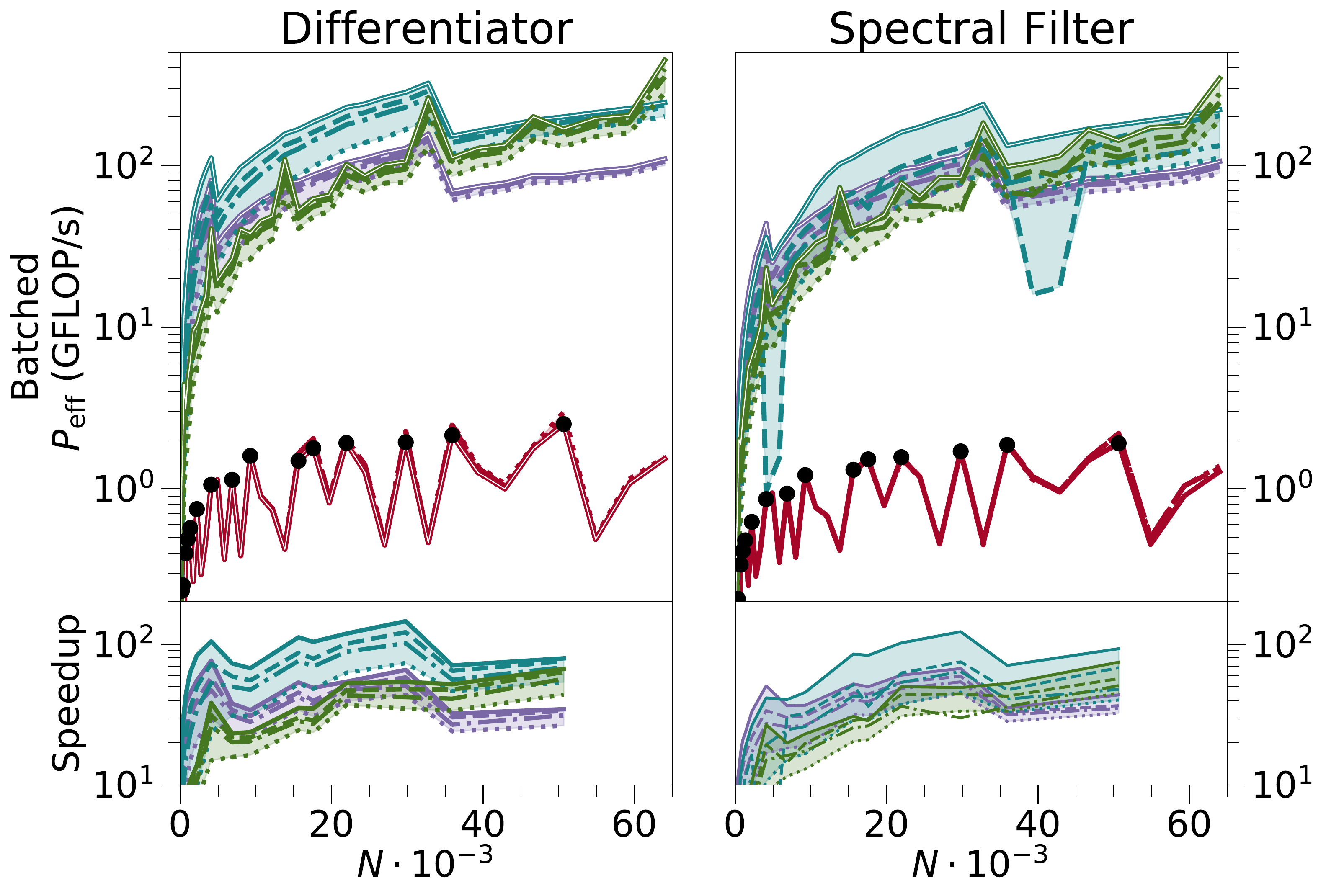}}
\centerline{\includegraphics[width=0.9\textwidth]{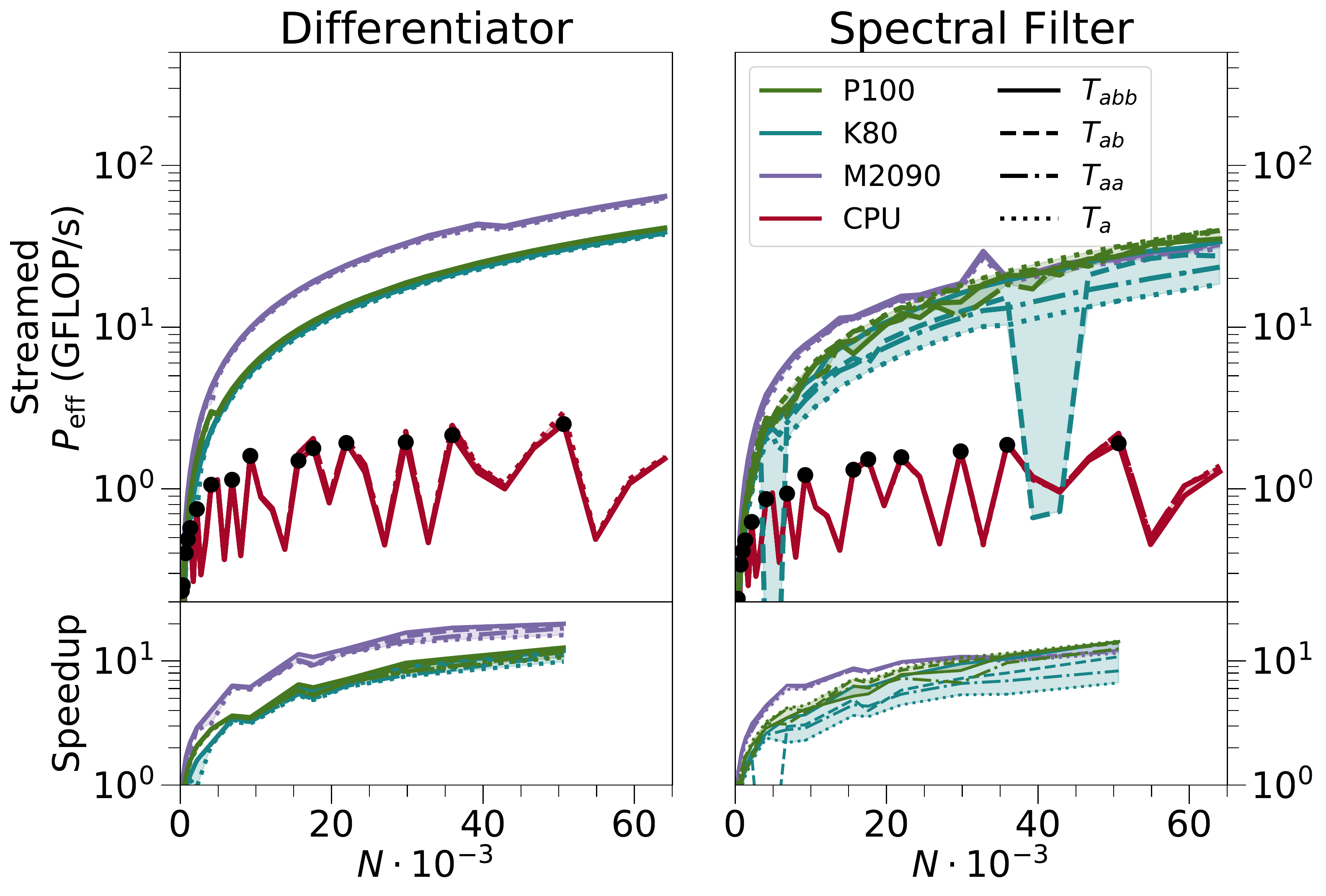}}
\caption{
  Effective processing power $P_\mathrm{eff}$ and speedups vs. one CPU core for 
   the matrix multiply differentiator (left)
  and spectral filter (right) acting on an $\mathcal{I}1 \otimes \mathcal{I}1 \otimes \mathcal{I}1$ topology. Input tensor structures differ by linestyle, while the 
  devices of Table \ref{tab:GPUtable} differ by colour. 
  The CPU algorithm exhibits sharply
  gridsize-dependent performance, and we compute speedups only at peaks, marked
  with black circles. 
  In the top (bottom) panels, we use the batched (streamed) API, which 
  in this case performs better (worse).
 }
\label{fig:diffI1batched}
\end{figure}

We now turn our attention to spectral filtering, and return focus
initially to $\mathcal{I}1 \otimes \mathcal{I}1 \otimes \mathcal{I}1$
topologies. 
Spectral filtering of 1D basis functions is similar to differentiation,
the only difference being the specific form of the transformation matrix. In 
Figure \ref{fig:diffI1batched} we thus show the performance of both the differentiator
and the spectral filter operating on an $\mathcal{I}1 \otimes \mathcal{I}1 \otimes \mathcal{I}1$ topology.
Comparing the performance of the spectral filter with that of the differentiator, 
we see near-identical behaviour on the CPU. On the GPU we get qualitatively similar
but somewhat worse performance from the spectral filter.
This is due to the extra cost in the latter case of looking up the cached transform 
matrices. Since the differentiator always implements the same transformation,
we can store the relevant matrices as private members of a differentiator C++
class. For the spectral filter, there are very many possible transformations,
which necessitates a more complicated caching strategy. While the performance 
difference is likely unimportant in practice,
the lookup could probably be substantially optimized if necessary.

%\begin{figure}[t]
%\centering
%\centerline{\includegraphics[width=1\textwidth]{plots/singlebh/difffiltI1streamed.pdf}}
%\caption{
  %Effective processing power $P_\mathrm{eff}$ and speedups vs. one CPU core for 
   %the matrix multiply differentiator (left)
  %and spectral filter (right) acting on an $\mathcal{I}1 \otimes \mathcal{I}1 \otimes \mathcal{I}1$ topology, plotted
  %against simulation gridsize. Formatting is identical to that of Figure \ref{fig:diffI1batched}, but these 
  %benchmarks are from the streamed API, which gives worse performance in this case. In particular,
  %the y-axis scale for the $P_\mathrm{eff}$ (but not the speedup) plots are identical to those in
  %Figure \ref{fig:diffI1batched}.
  %}
%\label{fig:diffI1streamed}
%\end{figure}
The CPU curves in Figure \ref{fig:diffI1batched} are computed using the same operation count model as
we use in the GPU case, which is $O(N^4)$ in the linear gridsize $N$. However,
the CPU in practice uses an FFT on the transformed basis function, and so
its true scaling is, for favourable collocation point choices, $O(N^3 \log{N})$.
Because our model underestimates the true CPU FLOP count the CPU performance
curves on Figure \ref{fig:diffI1batched} can in principle exceed the CPU's theoretical
performance (c.f. Figure \ref{fig:difftheory}), although in this case they do not.
The scaling coefficients at lower gridsizes are better for matrix multiply, which is 
why the latter alogorithm can be favourable, especially given superior hardware.
Unfavourable collocation point choices can furthermore affect the true CPU FLOP count by
about an order of magnitude, causing the jagged
behaviour of the CPU curves in Figure \ref{fig:diffI1batched}. When computing
the speedup, we use only the ``peak" points (chosen by eye), and have 
plotted a linear interpolation between these.

The performance of the CPU algorithm is roughly independent of the number of
independent tensor components $N_e$. For the GPU algorithm we can get 
some dependence upon the latter. The individual matrix multiplication sizes 
are on the order of the linear gridsize, between around 
$10$ and $40$. Neither \texttt{cuBLAS} nor the GPU itself are very well
optimized for such small matrix multiplications, which cannot individually
utilize all the streaming multiprocessors of the device.
This likely accounts for the underperformance of the GPUs
compared to their theoretical processing powers. The reason for the 
performance dip on the K80 $T_{ab}$ curves around $N=40000$ is unclear.

We thus use one of two concurrency strategies that allow
multiple small kernels to exhibit some parallelism. The first 
strategy, called ``streamed", attempts to run the kernels concurrently
using CUDA streams. These are a CUDA API feature that allow kernels to be run
asynchronously with the CPU and with one another. This approach cannot achieve
concurrent execution for very small kernels for which the kernel launch overhead
of about 20 $\mu$s is an important expense, since only one kernel
can be prepared for launch at one time. 
Also, since the individual kernels have no knowledge of one another, $\mathtt{cuBLAS}$
must tune them as if they were to run synchronously, which may result in 
suboptimal tuning overall.

\setcounter{footnote}{0}
The second strategy, called ``batched", runs each separate matrix transformation
as a single call to the API function $\mathtt{cublasDgemmBatched}$. This function
performs an identical matrix multiplication on a series of matrices, given to 
the API as 
an array of pointers. Using it incurs some extra overhead, since this array must
be first copied to the GPU \footnote{There is another API function, 
  $\mathtt{cublasDgemmStridedBatched}$, which avoids this overhead by accepting
  a single pointer for each matrix along with a stride that determines where in
  GPU memory each new matrix begins. We are unable to use this function since
our Tensor elements are not respectively contiguous.}. The batched API can in many cases give superior
performance to streamed multiplications. Generically, it will be the better choice
for numerous multiplications on small kernels. 
In that case the batched
API can save on launch overhead, and may also make superior 
tuning choices since it is aware of the full operation. 

Sometimes, however,
the streamed strategy is favourable. 
It
is not easy to predict which will be which except by experiment. We have,
for example, performed benchmarks which show that for some matrix shapes the
batched strategy is a factor of 2-5 faster even when only a single (small) matrix
is being operated upon. In other cases, we have found that the batched API is
modestly superior on some cards, but that streamed calls are almost an order
of magnitude better on newer ones, presumably because of new GPU features
being exploited on newer cards. Because of this, we have experimented with both 
strategies in all our benchmarks. 

On $\mathcal{I}1 \otimes \mathcal{I}1 \otimes \mathcal{I}1$, 
the batched strategy consistently gives an improvement of about an order of 
magnitude, with larger tensors giving a greater advantage. This topology
involves very many individually tiny matrix multiplications ($N_e(2 + N)$ in total),
so this is perhaps to be expected. Especially when using the batched strategy, we
get very impressive speedups overall, of between 10 and 100X. This is despite
the observed performance being about an order of magnitude beneath our theoretical
prediction. It must be stressed
that our CPU benchmarks use only 1 CPU core, which has a fairly modest clock frequency
of 2.0 GHz. While realistic for $\mathtt{SpEC}$ this would 
in most circumstances be a very unusual comparison. 6 CPU cores running at 
3-4GHz might be more typical of modern hardware, which would give about an
order of magnitude speedup assuming linear scaling with parallelism (which 
$\mathtt{SpEC}$ cannot achieve).

\begin{figure}[t]
\centering
\centerline{\includegraphics[width=0.9\textwidth]{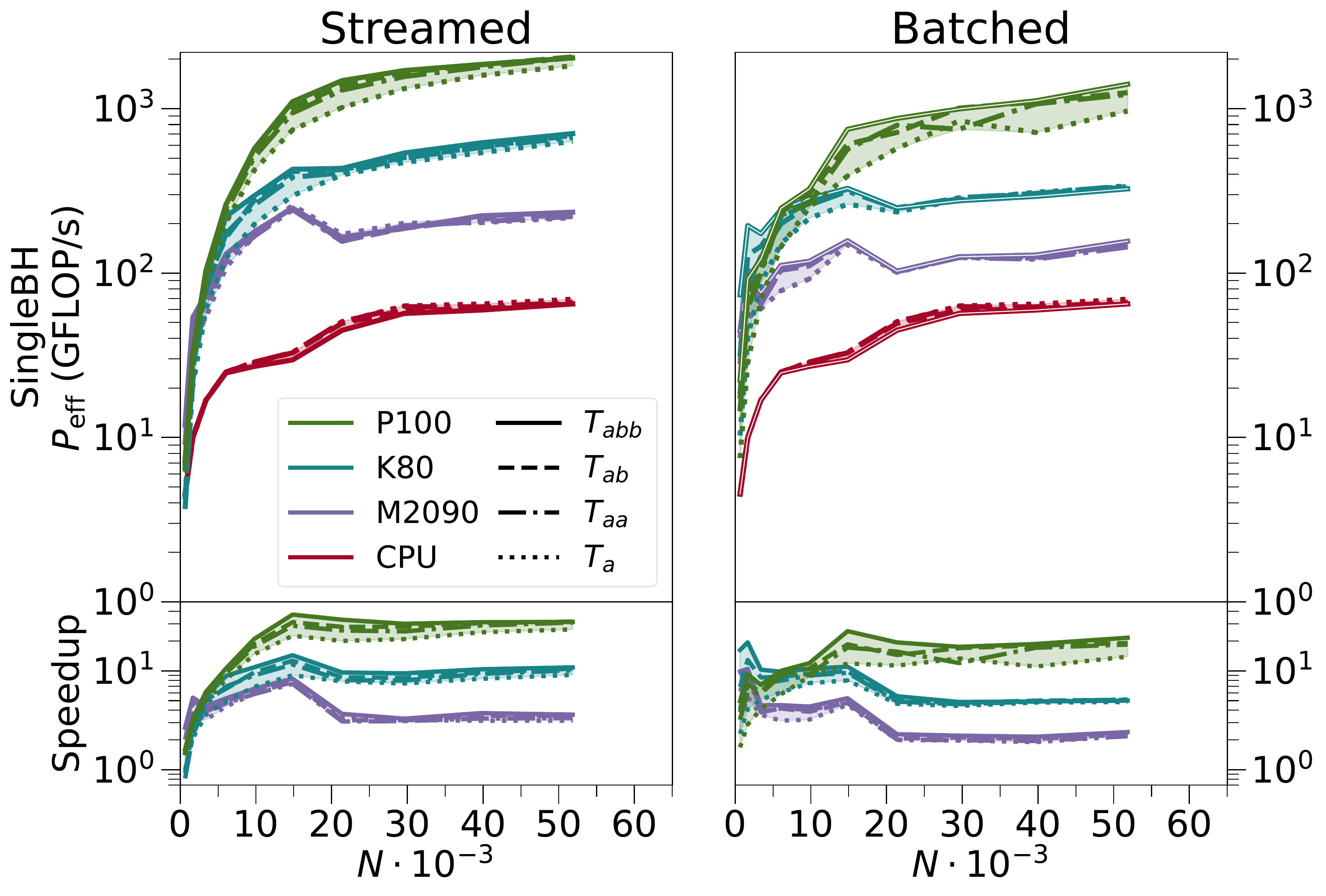}}
\caption{
  Effective processing power $P_\mathrm{eff}$ and speedups vs. one CPU core for the matrix multiply differentiator acting on a $\mathcal{I}1 \otimes \mathcal{S}2$ topology using the `SingleBH' gridpoint distribution, using the streamed (left) and batched (right)
  concurrency strategy. In terms of the resolution parameter $k$, `SingleBH' has $N_r=9+4k$,
  $N_\theta=6+2k$, and $N_\phi=2N_\theta$.
  Different linestyles indicate differing tensor structures as indicated in the legend, with a 
  colour fill between $T_{abb}$ and $T_{a}$ (performance of the intermediate structures is usually, but not always, bounded by these). 
}
\label{fig:diffS2sbh}
\end{figure}
\begin{figure}[t]
\centering
\centerline{\includegraphics[width=0.9\textwidth]{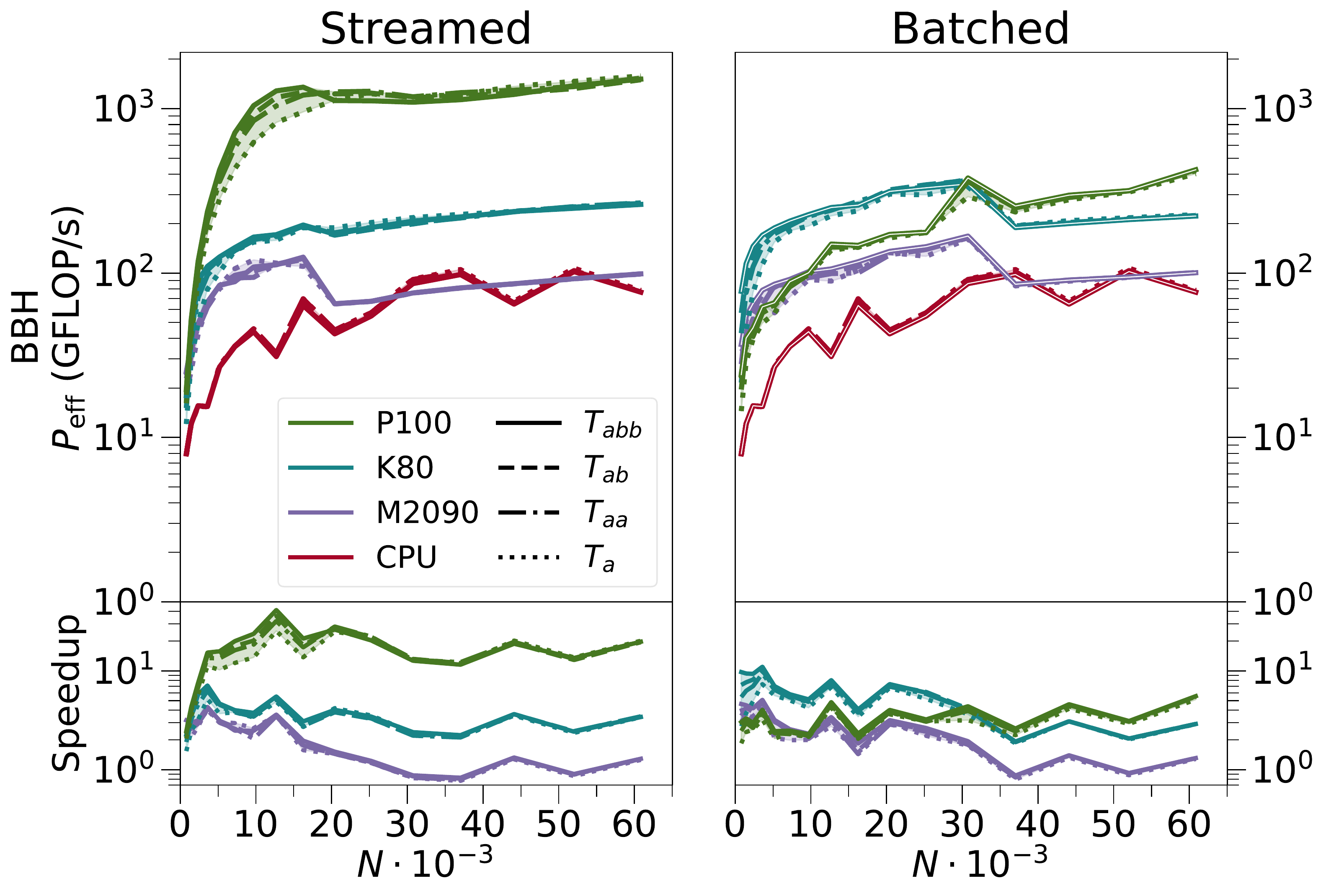}}
\caption{
 Effective processing power $P_\mathrm{eff}$ and speedups vs. one CPU core for the matrix multiply differentiator acting on a $\mathcal{I}1 \otimes \mathcal{S}2$ topology using the `BBH' gridpoint distribution, using the streamed (left) and batched (right)
  concurrency strategy. In terms of the resolution parameter $k$,  `BBH' has $N_r=4+k$, $N_\theta=7+2k$, and $N_\phi=2N_\theta$.
  Formatting and axis scales are identical to those in Figure \ref{fig:diffS2sbh}. 
}
\label{fig:diffS2bbh}
\end{figure}
We now turn to the results on spherical shells $\mathcal{I}1 \otimes \mathcal{S}2$,
where a more complex picture will emerge than for $\mathcal{I}1 \otimes \mathcal{I}1 \otimes \mathcal{I}1$. We first discuss the differentiator, results from which are 
summarized in Figures \ref{fig:diffS2sbh}-\ref{fig:diffS2bbh}. Especially for larger gridsizes and especially
on the P100 our GPU performance is quite comparable to the predicted peak
performance shown in the lower panels of
Figure \ref{fig:difftheory}. The CPU performance, on the other hand, 
exceeds both this prediction, and the CPU's theoretical processing power. 
The
expense of the transform is dominated by the angular sector in both cases. 
On the CPU, the $\phi$ transform is done with an FFT, and the $\theta$ by 
a matrix multiply, yielding 
$N_r (N_\theta N_\phi)^3 \log{(N_\theta N_\phi)}$ scaling, compared to the
$N_r (N_\theta N_\phi)^4$ scaling of our model. 
This gives a ratio $N_\theta N_\phi / \log{(N_\theta N_\phi)}$, which is a larger
factor than for $\mathcal{I}1 \otimes \mathcal{I}1 \otimes \mathcal{I}1$, since
$N_\theta N_\phi$ is larger. This is particularly true for the BBH case, 
explaining the improved CPU performance
of BBH vs. SingleBH.

For the SingleBH grid we achieve an appreciable speedup of between $5$ and $30$X 
throughout. Performance is consistently better for the streamed strategy in this 
case, perhaps because the individual angular multiplications are now large enough
that $\mathtt{cuBLAS}$ can make effective tuning choices. For BBH, where the
spectral algorithm gives the largest advantage, the GPU advantage is more modest
and the CPU actually
exceeds the M2090 performance in some cases. This is unfortunately the more realistic
gridsize choice for production simulations. 

The batched vs. streamed picture is here much less 
clear than it was for $\mathcal{I}1 \otimes \mathcal{I}1 \otimes \mathcal{I}1$. 
On the P100 the streamed strategy is greatly advantageous, whereas
batched is modestly superior on the other cards. GPU performance seems in most
cases to scale up to some kind of threshold, after which point there is a sharp
dip and a new slow scaling upwards (for example at around gridsize 30000 for the
BBH batched K80 and M090 runs, or 15000 on SingleBH). This may be due to $\mathtt{cuBLAS}$
switching here to a new kernel optimized for multiplications with a large shared dimension.
In that case, kernels need to read much more data than they will end up writing
to global memory, which limits parallelism.

We now turn to spectral filtering on $\mathcal{I}1 \otimes \mathcal{S}2$ shells.
The radial filter, along $\mathcal{I}1$, is handled
in the same way as the $x$-dimension in $\mathcal{I}1 \otimes \mathcal{I}1 \otimes \mathcal{I}1$. In our benchmarks, as in production runs, we do not include a filter along the
$r$-axis.

For spectral filtering of 2D
basis functions, the filtering transformation will normally couple
together elements from different components of a tensor. 
This necessitates a 
different approach, especially since the relevant coupling will be
very sparse. 
For filtering on  $\mathcal{I}1 \otimes \mathcal{S}2$ we therefore break the operation into three steps. During
``PhysToSpec" (``SpecToPhys"), each of the input tensor $U_{ijk...}$'s $N_e$ independent 
components are separately transformed as in Eqs. \eqref{eq:PhysToSpec}
and \eqref{eq:SpecToPhys}. In physical space, each tensor component is viewed 
as a $(N_r, (N_\theta N_\phi))$ matrix, and the spectral transform matrix
has dimensions $((N_\theta N_\phi), N_s)$, where the spectral dimension $N_s$
(very roughly $N_\theta N_\phi / 2$) is the number of coefficients of 
the spectral representation. Thus,
\begin{align}
  F &= N_e N_r N_s (2N_\theta N_\phi - 1), \\
  M &= N_e (N_r N_\theta N_\phi + N_r N_s + N_\theta N_\phi N_s).
\end{align}
The ``SpecToPhys" transformation simply swaps
$N_\theta N_\phi$ with $N_s$ in the above. 

We choose BLAS parameters such that the PhysToSpec transform maps the different
tensor components into a single contiguous $N_r N_e N_s$ array, which we view
as an $(N_e N_s, N_r)$ matrix. The filter is then implemented by multiplying
with an $(N_e N_s, N_e N_s)$ transform matrix which couples the $N_e$ distinct
tensor components. Typically, only
about 10\% or fewer of the entries in this matrix are nonzero, and so it becomes
worthwhile to store it in a sparse format. We use the CSR format \cite{cuSPARSE} because it is
fairly simple and well-supported by the NVIDIA sparse algebra package 
$\mathtt{cuSPARSE}$.

The complexity of the filtering step depends somewhat sensitively upon the actual
structure of the sparse filtering matrix and upon the details of the matrix 
multiplication algorithm. The sparsity of the filtering matrix will in turn depend
on what filter is being applied, so we profile using two different such functions,
which we call Heaviside (a Heaviside filter) and ExpCheb (an exponential Chebyshev function). 
We use the $\mathtt{cuSPARSE}$ algorithm 
$\mathtt{dcsrmm2}$. The \texttt{cuSPARSE} documentation \cite{cuSPARSE} 
describes this algorithm as memory bound, with an approximate complexity of
$N_eN_s[s N_eN_s(N_eN_s+1) + 2 N_r]$. Here the sparsity
factor $s = Nnz / (N_e^2 N_s^2)$, while $Nnz$ is
the number of nonzero entries in the sparse matrix. 

\begin{figure}
\centering
\centerline{\includegraphics[width=0.9\textwidth]{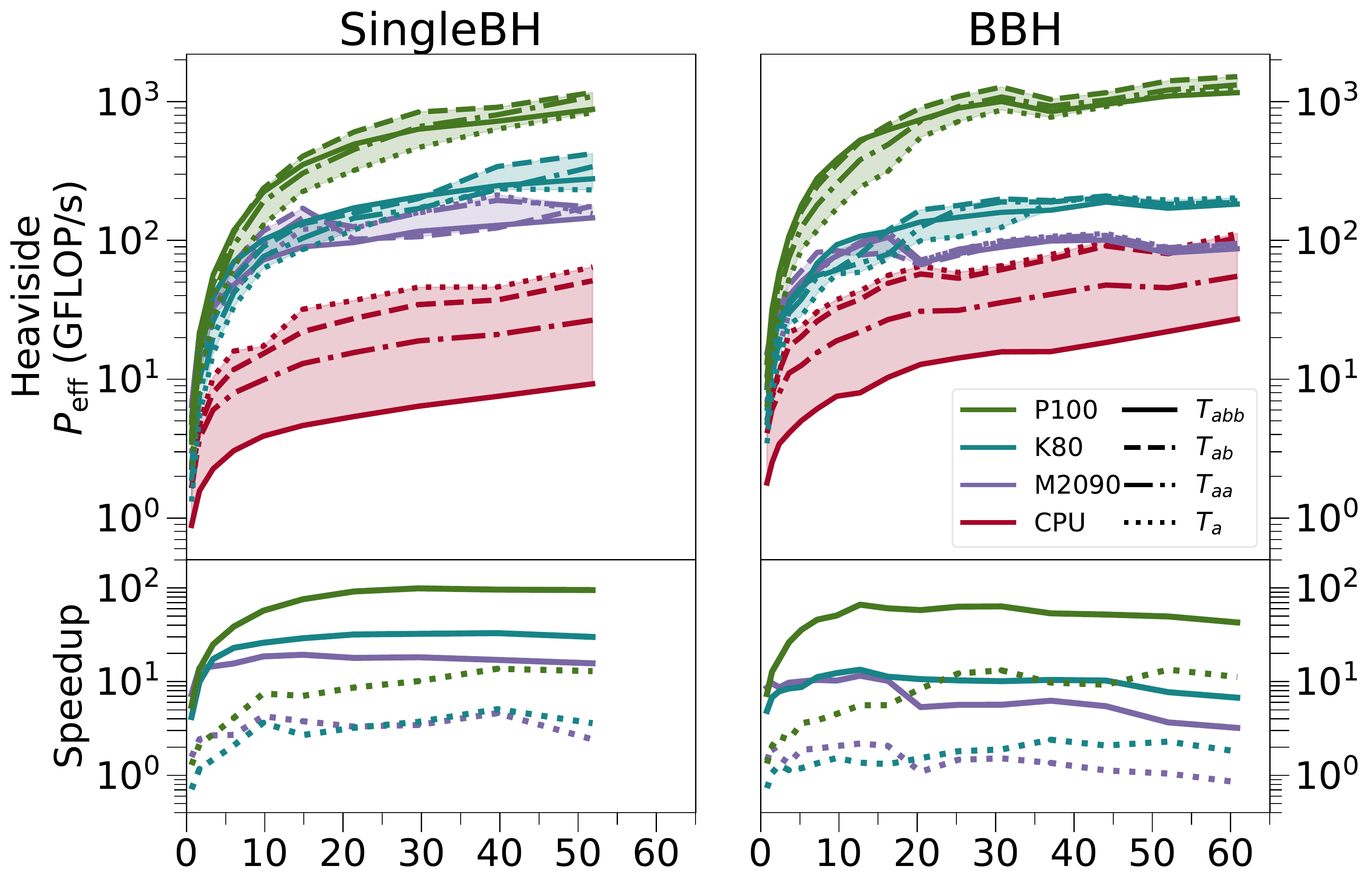}}
\centerline{\includegraphics[width=0.9\textwidth]{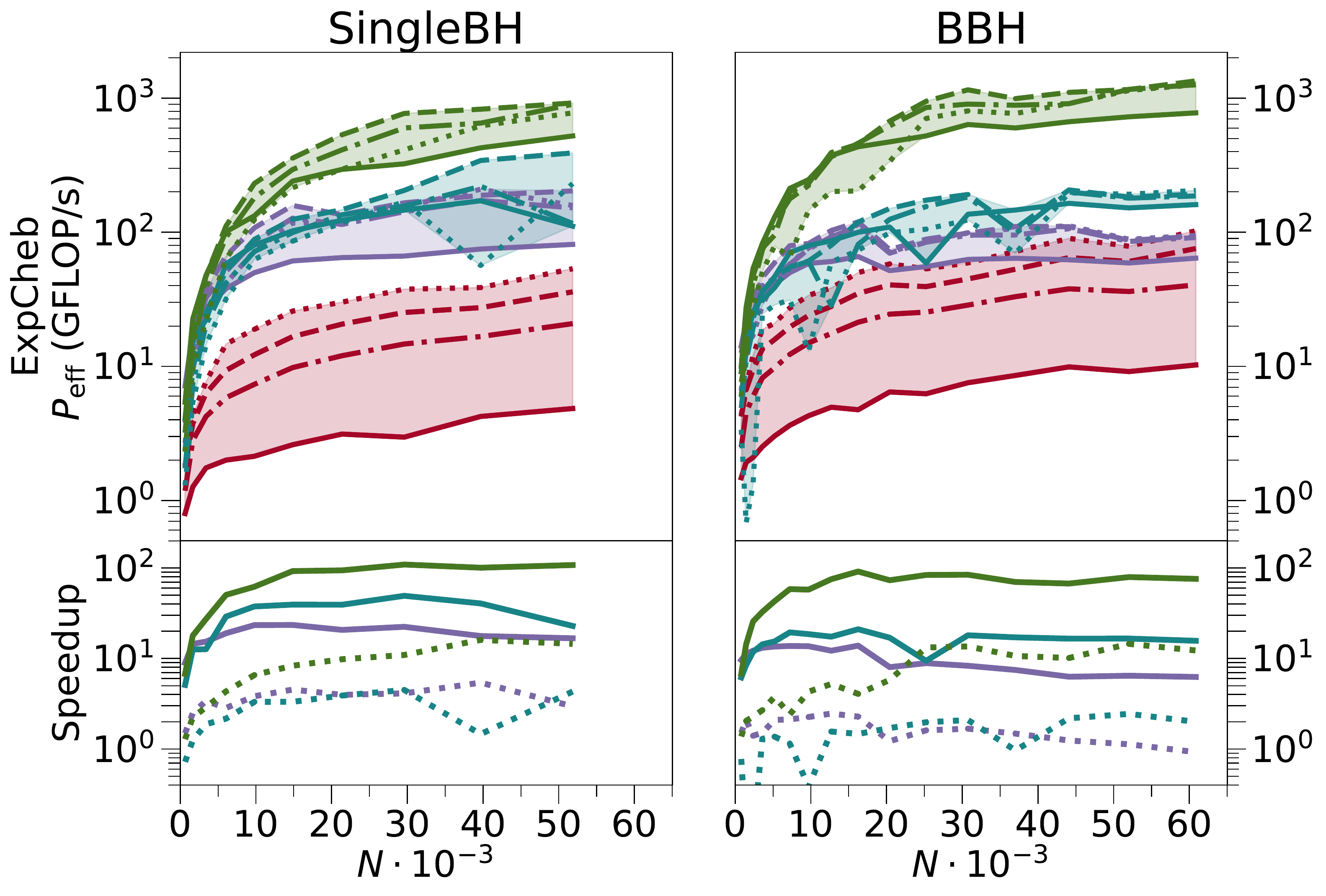}}
\caption{
  Effective processing power $P_\mathrm{eff}$ and speedups vs. one CPU core 
  for the matrix multiply spectral filter acting on an $\mathcal{I}1\otimes\mathcal{S}2$ 
  topology for `SingleBH'
  (left) and `BBH' (right) gridpoint distributions. We study two filter functions,
  Heaviside (top) and ExpCheb (bottom). Speedups are shown only for $T_\mathrm{abb}$ 
  and $T_\mathrm{a}$. Line colours (and fill) distinguish between processors,
  and line styles between tensor structures. Here we only profile
  the streamed API, which generally performs better.
}
\label{fig:sfS2}
\end{figure}

The benchmarks in this case are illustrated by Figure \ref{fig:sfS2}. GPU
performance is dominated by the (dense) spectral transform multiplications, and
so the results are comparable to, but worse than,  those of the differentiator acting on 
$\mathcal{I}1 \otimes \mathcal{S}2$, shown in Figures \ref{fig:diffS2sbh}-\ref{fig:diffS2bbh}. 
The worse performance is due to the
larger number of operations, the extra time needed for matrix lookup, and the
more asymmetrical matrix dimensions. Speedups, however, are in many cases higher
for the spectral filter, 
because in that case the CPU performance is much more sensitive to the 
number of independent components of the input tensor. Curiously, the CPU processes
$T_{ab}$ considerably more efficiently than $T_{aa}$, even though the latter
has fewer components. As for the differentiator,
the streamed concurrency strategy gives somewhat better results overall for
SingleBH. For BBH, the batched API is marginally superior except on the P100,
where the streamed strategy outperforms by about a factor of 2 (c.f. the hollow lines
in the lower panels of Figures \ref{fig:stackplotssbh} and \ref{fig:stackplotsbbh}).

\subsection{$\mathtt{DataMesh}$ Operations, Apply BCs, GHEqns}
Apart from the individually-significant operations described above, $\mathtt{SpEC}$ contains
numerous operations on our array class, $\mathtt{DataMesh}$, which are distributed
too widely throughout the code to port individually. To deal with these we
use our automatic porting system, \texttt{TLoops}. While \texttt{TLoops} is a
separate subject in itself, we briefly describe it here to keep this paper self-contained.

\texttt{TLoops} furnishes a set of C++ classes to represent arrays, tensors, indices
over tensors, and operations between tensors. These classes are based on templates,
which recursively iterate at compile time to form unique types for each tensor
manipulation written in the $\mathtt{SpEC}$ code. After compilation a separate executable
can be used to generate valid CUDA code for each unique operation. This can
then be linked back to a separate checkout of $\mathtt{SpEC}$. In this way all manipulations
of $\mathtt{DataMesh}$ throughout the code can be ported at once.

If no changes are made to the code at all, $\mathtt{DataMesh}$ operations ported automatically 
with \texttt{TLoops} will typically show only
a modest speedup or even a slowdown at lower gridsizes, due to large amounts 
of launch overhead incurred by loops over kernel launches\footnote{Concurrent
  kernel execution using CUDA streams does not help in the case where launch
overhead is more expensive than the kernel itself.}. Even in the case of a small
slowdown, however, the automatic porting yields a net benefit since it avoids 
numerous CPU-GPU synchronizations that would otherwise
occur around the explicitly ported modules.

Much of the slowdown comes from the operations collected as ApplyBCs. These
functions operate mostly on angular slices at the boundaries of the domains, 
which have about an order of magnitude fewer points than do the full
three-dimensional volume arrays: a shell with $(N_r, N_\theta, N_\phi)$ gridpoints
has boundaries with only $(N_\theta, N_\phi)$ gridpoints, and $N_r$ is typically
around $10$. Such operations can be a considerable bottleneck
for two reasons. First, the code that extracts and inserts these two-dimensional
slices out of and into the volume involves an unavoidably strided data access
that is very inefficient to port on the GPU. It is nevertheless best to do so
in order to avoid extra synchronizations. Second, the ApplyBCs operations
are simply very small and very numerous. Launch overhead impairs their 
performance severely.

We deal with this by leaving boundary data on the CPU throughout. \texttt{TLoops} 
expressions check the dimension of the relevant $\mathtt{DataMesh}$es, and execute on 
the GPU only for dimension 3 or higher. The $\mathtt{DataMesh}$ copy constructor also
transfers data on the host (rather than the device) for dimension 1 or 2, 
unless the data is on the 
device already (for dimension 3 we always synchronize with the GPU and copy there).
\texttt{TLoops} still somewhat impairs the performance of the ApplyBCs operations
since the boundary arrays must be copied to the GPU whenever they are to be
extracted from or inserted to the volume. But the net effect is a speedup.

Launch overhead can be mitigated even further by operating on whole tensors
with a single kernel. Code must be modified to do this, so it is not practical
to do throughout, but it does provide a very simple and convenient porting
strategy for complicated operations. We have used this strategy to port the 
GHEqns, which solve the Einstein equations. This allows for about a 10X
speedup at realistic gridsize without writing any explicit CUDA code (c.f. Section 4.4).

\section{Benchmarks of Overall Code}
\label{OverallBenchmarks}
\begin{figure}
\begin{center}
\includegraphics[width=\textwidth]{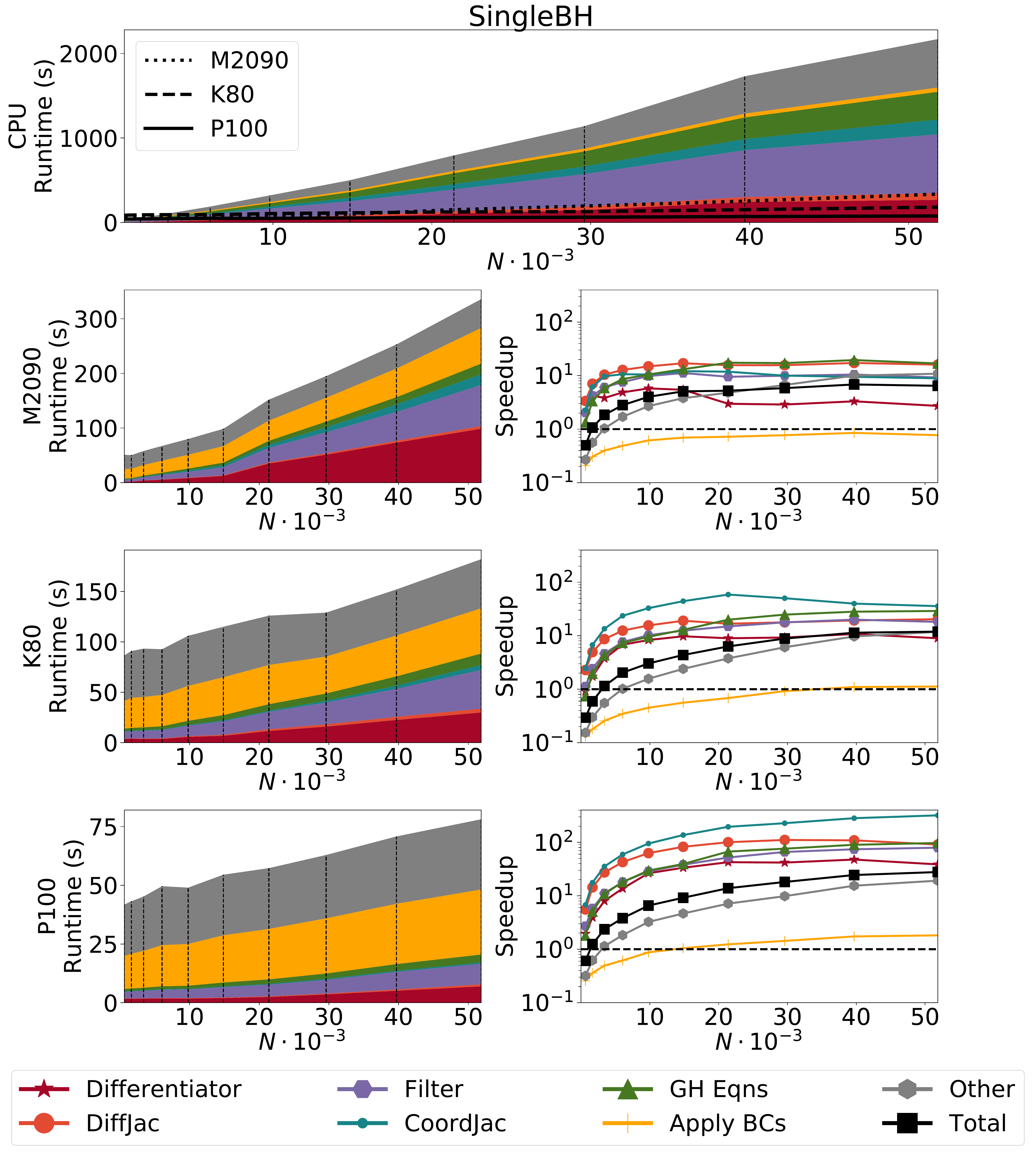}
\caption{$\mathtt{SpEC}$'s overall performance broken down by module using the
  SingleBH gridpoint distribution. On the top and left, we show ``stack" plots
  for each benchmarked device. The height of each slice is the runtime of its 
  module, and the total height is the overall runtime. 
  Dotted vertical lines mark the profiled resolutions.
  On the CPU stack (top), we overlay the GPU runtimes as black lines.
  The grey
  stack (`Other') represents all modules not otherwise included. On the right,
  we show the speedup of each module, with identical colouring as on the left,
  and the total speedup in black.}
 \label{fig:stackplotssbh}
  
\end{center}
\end{figure}

\begin{figure}
\begin{center}
\includegraphics[width=\textwidth]{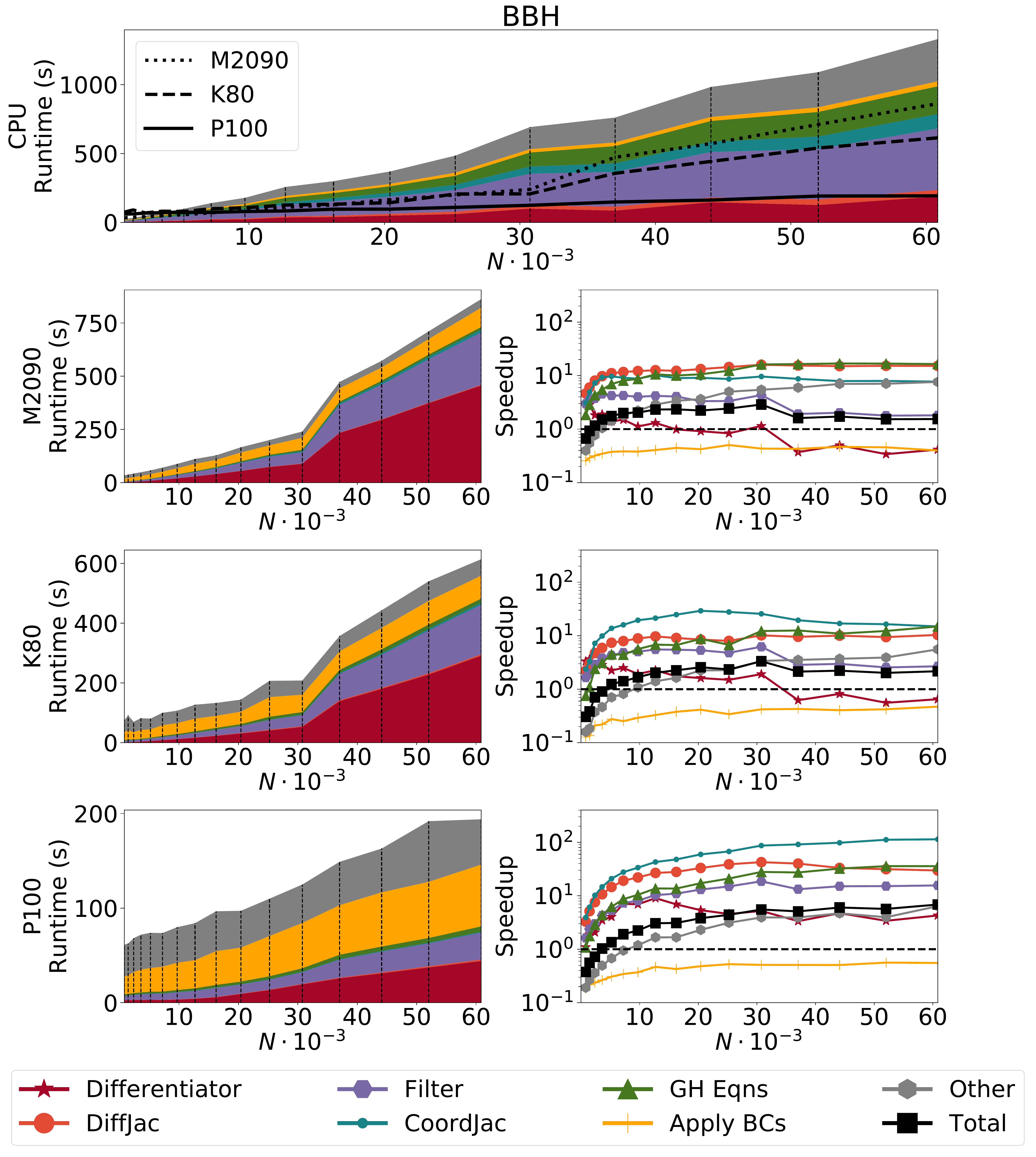}
\caption{$\mathtt{SpEC}$'s overall performance broken down by module using the
  BBH gridpoint distribution. The plots are formatted in the same way 
as Figure \ref{fig:stackplotssbh}.}
\label{fig:stackplotsbbh}
\end{center}
\end{figure}

Figures $\ref{fig:stackplotssbh}$ and $\ref{fig:stackplotsbbh}$, finally,
summarize our entire GPU porting results. These figures show
benchmarks from runs of $\mathtt{SpEC}$ upon isolated black hole
test cases. Shown are two gridsizes, SingleBH and BBH, identical to the 
eponymous gridsizes used in the performance analysis of the differentiation and
the spectral filter in \ref{spectral}. Compared to the SingleBH tests, the BBH 
tests have a relatively
larger angular resolution at constant gridsize. We ran each benchmark five times 
for 110 timesteps, and collected results between timesteps
5 and 105. As in the previous benchmarks, the plotted results are the 
median times over the five runs. 

These single black hole runs evolve a stationary, single black hole with a spin of
$0.5$ using the generalized harmonic equations. Surrounding
the black hole are two $\mathcal{I}1 \otimes \mathcal{S}2$ subdomains with identical resolutions.
Apart from being numerically simpler, the isolated case differs from a full binary black
hole
simulation in several ways. A binary black hole simulation would have a much more diverse set
of domains, and would involve AMR, which we have not considered here. Binary evolutions
also involve interpolations and computations on the apparent horizons of the black
holes, which can be importantly expensive. Finally, binary evolutions use MPI, 
which assigns each individual
domain to a different CPU core. In our full port, each would instead be
assigned to a separate GPU. However, in these tests, we do our computations on the
two respective spheres in serial.

The GPU performance is especially strong in the (less realistic) SingleBH gridpoint 
distribution, 
which has more points in the radial direction at a given gridsize. 
Generally we see comparable performance for both distributions as expected from our individual benchmarks
in the per-module speedups, although the differentiator on the BBH distribution performs
somewhat worse than expected. The GHEqns, which are ported automatically
using \texttt{TLoops},  show particularly 
strong performance - a 100X speedup on the P100 - especially given that no 
algorithmic redesign was required here.

The differentiator, and to a lesser extent the filter, remains important to the 
overall wallclock time, particularly for P100. Indeed, the overall speedup
is limited by this module, especially on the M2090. Porting the spectral algorithm
explicitly on spheres may be worthwhile.

Much of the speedup is limited by the ApplyBCs and the Other operations. The latter
are mostly a large mass of tensor manipulations which have not been explicitly replaced with
\texttt{TLoops} expressions. We expect a further speedup of these modules by 2-15X could be achieved 
by doing so, due to both savings in launch overhead and extra parallelism over
tensor structure.

The ApplyBCs operations present a more serious challenge. These operations are 
on arrays of dimension 2, which we have purposely left on the CPU throughout.
The performance of these operations is thus limited in principle by the CPU
performance\footnote{We get a slight speedup at times on the K80 and P100 operations because
these GPUs are driven by POWER8 CPUs with a higher clock frequency than we used
for our CPU benchmarks}. It will usually be somewhat worse than this, because these low-dimension
arrays arise as slices of higher-dimensional data living on the CPU, and extracting
these slices requires a GPU-CPU memory transfer. The arrays in question are 
so small that kernel launch overhead is the dominant expense if they are kept on the 
GPU throughout, to a sufficient extent that the memory copies are still cheaper 
overall. However, porting to \texttt{TLoops} may mitigate this. 

Parallelization of a particular domain across multiple GPUs would not be worthwhile,
as our essential problems throughout have been code complexity and the small
size of our operations compared to those for which the GPU is optimized. However,
multiple GPUs can still be leveraged by assigning one domain to each, which should
give roughly linear scaling.

\section{Conclusions}
We have performed a CPU to GPU port of the portions of the numerical relativity
code $\mathtt{SpEC}$ relevant to single black hole simulations. Our combined
strategy of explicit porting for the Jacobian multiplications, semi-automatic
porting for spectral operations that can be written as matrix transformations,
and completely automatic porting for the many scattered tensor operations throughout
the code gives comparable to peak performance for many of these modules, module-to-module
speedups compared to one CPU core ranging from 10 to 100X, and overall speedups
of between 2-10X. Due to its reliance on prepackaged libraries, our port also 
generally shows improved performance when run on newer hardware without requiring extra tuning.

The next step in GPU-porting will be to completely port the code so
that the accelerated version can be used for binary black hole
simulations.  Binary black hole runs differ from the single black hole
case in three main ways. First, they involve a larger and more diverse
set of computational domains than the two spherical shells considered
in the single black hole case. Second, those domains must
communicate in parallel using MPI.  Third, production runs involve
apparent--horizon finding.  Let us comment on each of these issues in turn:

The specific individual domains involved in a binary black hole run are spherical 
shells ($\mathcal{I}1 \otimes \mathcal{S}2$), 
cubes ($\mathcal{I}1\otimes\mathcal{I}1\otimes\mathcal{I}1$),
and cylindrical shells ($\mathcal{I}1\otimes\mathcal{B}2$). Since our port
relies on ``black box" linear transformations, the most important feature of 
these domains from our perspective is their spectral dimension; i.e. 
$\mathcal{S}2$ (dimension 2) vs $\mathcal{I}1$ (dimension 1). As documented
in Section 3 our port performs best relative to the CPU spectral algorithms
in the dimension 1 case. Our single black hole runs, however, were profiled
on dimension 2 spherical shells, which apart from being dimension 2 make
use (on the CPU) of heavily-optimized FORTRAN modules for the efficient 
manipulation of spherical harmonics. Internal to each domain, we can therefore
expect performance during production runs at least equal to, and probably better
than, during the single black hole simulations considered in the present work.

  Moving from one subdomain to many will change the relative importance of
  boundary condition operations (`ApplyBCs') compared to volume data
  processing.  Since the ApplyBCs modules were problematic for our
  port, this could conceivably impair scaling to larger
  domains. However, we in fact expect this scaling to be favourable.
  SpEC runs use ApplyBC in three distinct cases: (i) at `internal'
  boundaries between neighboring domains, the characteristic variables
  are computed and used as boundary conditions on the neighboring
  subdomain; (ii) at the external outermost boundary, rather complex
  constraint-preserving boundary conditions are
  applied~\cite{Szilagyi:2009qz}; and (iii) at BH excision boundaries,
  no boundary conditions are applied at all.  The SingleBH test-case
  presented here uses two domains, leading to two internal
  boundaries and one external boundary with the constraint
  preserving boundary conditions (plus one `no-op' at the
  inner excision boundary).  
  
  For a full binary black hole simulation,
  there will still be only \emph{one} outer boundary where the
  expensive (ii) boundary condition is applied.  The number of inner
  boundaries (i) will grow in proportion to the number of domains
  (plus there will be two `no-op' black hole excision boundaries).
  Therefore, we expect that the computational cost of ApplyBC relative
  to the volume operations will decrease as more domains are added,
  because of the smaller fraction of boundaries with the expensive
  (ii) boundary conditions.
%conditions between
%the two spherical shells, and one of the `constraint' boundary conditions
%at the edge of the entire domain (at the edge of the outermost shell). The addition
%of more subdomains would require an extra application of internal boundary
%conditions per new subdomain, but the constraint conditions are only ever
%applied once. However, the most problematic modules in our profiled runs
%were constraint conditions.

There is also the practical matter of interfacing MPI with a multi-GPU code.
In our case this is quite simple
since our model of parallelism has each GPU operating on data-independent
domains. The business of `load-balancing' between independent processors
is already handled by mature SpEC code.

The effect of apparent horizon processing on performance is harder to predict.
The apparent horizon will normally not be confined to a particular simulation
domain. Finding it thus requires data from multiple domains, and thus
entails addition CPU-GPU communication steps. After finding the horizon,
data must be interpolated upon it, an operation which can be cast as another
linear transformation. Without further study, it is difficult to speculate
on the overall impact this will have.

  A recurring question in GPU computing is whether the quite
  substantial effort of a GPU-port is worthwhile at the end.
  Reorganizing the structure of filtering and derivative-computation
  toward fewer, larger BLAS calls was a requirement for the CUDA port
  presented here.  This same reorganization also had the side-benefit
  of speeding up the CPU version of these modules by a factor of
  $\sim\!\!2$, and has thus benefited SpEC in general.  An alternative
  approach to GPU porting would be to implement efficient
  multi-threaded processing across CPU-cores, using e.g. OpenMP.  The
  idea is to spread each domain over multiple compute-cores, possibly
  across an entire compute node, and so enable a different layer of
  parallelism.  This would still entail separate coding
  efforts for each major part of the code which we had to deal with in
  the present work, among them filtering, derivatives, general
  computations.  Past efforts within the SXS collaboration to utilize
  OpenMP to speed up tensor-computations like \texttt{GHEqns} have
  not led to a speed-up. Especially given the new opportunities
  afforded by the automatic code generation in TLoops, it would be
  worthwhile to renew such efforts.

\section{Acknowledgments}
We thank Nils Deppe and Mark Scheel for helpful discussions. 
Calculations were performed with the {\tt SpEC}-code~\cite{SpEC}. 
We gratefully acknowledge support from NSERC of Canada, form the Canada
Research Chairs Program, and from the Canadian Institute for Advanced Research.
Some calculations were performed at the SciNet HPC Consortium~\cite{scinet}.
SciNet is funded by: the Canada Foundation for Innovation (CF) under the auspices of
Compute Canada; the Government of Ontario; Ontario Research Fund (ORF) -- Research Excellence;
and the University of Toronto.

%%%%%%%%%%%%%%%%%%%%%%%%%%%%%%%%%%%%%%%%%%%%%%%%%%%%%%%%%%%%%%%%%%%%%%%%%%%%%%%
\section*{References}
%%%%%%%%%%%%%%%%%%%%%%%%%%%%%%%%%%%%%%%%%%%%%%%%%%%%%%%%%%%%%%%%%%%%%%%%%%%%%%%
\bibliographystyle{iopart-num}
\bibliography{sbh}

\end{document}